\documentclass[aps,prl,amsmath,amssymb,superscriptaddress,showkeys,showpacs,twocolumn,floatfix]{revtex4-1}

\immediate\write18{texcount -inc -sum 20260820_main.tex > wordcount.txt}

\usepackage{amsmath,amssymb,mathtools}
\usepackage{mathrsfs} 
\usepackage{graphicx,color}
\usepackage{float}
\usepackage{appendix}
\usepackage{hyperref}
\usepackage{verbatim}
\usepackage{comment}
\usepackage{subcaption}
\usepackage{bm}
\usepackage{xcolor}
\usepackage[justification=raggedright,singlelinecheck=false]{caption}

\usepackage[utf8]{inputenc}
\usepackage[english]{babel}
\usepackage{booktabs}
\usepackage[normalem]{ulem}

\begin{document}

\title{Solving the Domain Wall Problem: Primordial Magnetic Fields as a Dynamical Solution}

\author{Zizhuo Zhao}
\affiliation{
~International Centre for Theoretical Physics Asia-Pacific (ICTP-AP), University of Chinese Academy of Sciences (UCAS), Beijing, China.
}
\affiliation{
~Taiji Laboratory for Gravitational Wave Universe (Beijing/Hangzhou), University of Chinese Academy of Sciences (UCAS), Beijing, China.}

\author{Yuefeng Di}
\affiliation{
Institute of Theoretical Physics,
Chinese Academy of Sciences, Beijing 100190, China}
\affiliation{School of Physical Sciences, University of Chinese Academy of Sciences (UCAS), Beijing 100049, China}

\author{Ligong Bian\thanks{Corresponding Author.}}
\email{lgbycl@cqu.edu.cn}
\affiliation{~Department of Physics and Chongqing Key Laboratory for Strongly Coupled Physics, Chongqing University, Chongqing 401331, P. R. China}

\author{Jing Shu\thanks{Corresponding Author.}}
\email{jshu@pku.edu.cn}
\affiliation{~School of Physics and State Key Laboratory of Nuclear Physics and Technology, Peking University, Beijing 100871, China}
\affiliation{
~Center for High Energy Physics, Peking University, Beijing 100871, China}
\affiliation{
~Beijing Laser Acceleration Innovation Center, Huairou, Beijing, 101400, China}

\begin{abstract}
To prevent overclosing the Universe, stable domain walls must decay—a problem typically addressed by breaking the Peccei-Quinn symmetry with a bias term. In this Letter, we present a novel mechanism in which a helical primordial magnetic field provides a spatially inhomogeneous Chern-Simons source that drives the axion field. We validate this mechanism through three-dimensional lattice simulations of the evolving domain-wall network, demonstrating its efficacy in inducing decay. Our quantitative analysis reveals that the correlation length of the primordial magnetic field is a crucial determinant of the decay rate of the network and the properties of the emitted axions and gravitational waves. In particular, the axion emitted from the decaying domain-wall network can contribute to the dark matter abundance, and the associated gravitational-wave signal lies in the frequency range probed by pulsar timing array experiments.

\end{abstract} 
\maketitle

\noindent{\it \bfseries Introduction.}
The QCD axion was originally proposed to solve the strong CP problem \cite{PQ_1,PQ_2,PQ_3,PQ_4}, and axion-like particles serve as compelling dark matter (DM) candidates \cite{axion_DM_1, axion_DM_2}, raising many terrestrial and astrophysical probe interests~\cite{Irastorza:2018dyq,RevModPhys.93.015004}. After Peccei–Quinn (PQ) symmetry breaking in post-inflationary scenarios, the ultraviolet complete model with the standard model fermions, which acquire PQ charge, predicts the formation of axion domain wall (DW) networks via non-perturbative effects after the QCD phase transition~\cite{Dine:1981rt,Zhitnitsky:1980tq}. These networks may eventually dominate the cosmic energy density, giving rise to the cosmological DW problem \cite{DW_dominant_1, DW_dominant_2}.

Various solutions for the domain wall problem have been proposed, such as introducing a bias term between different vacua in the potential \cite{bias_1, bias_2, bias_3, bias_4}, the Lazarides–Shafi (LS) mechanism \cite{LS_1, LS_2, LS_3}, the introduction of a light axion-like particle (ALP) coupled to gluons \cite{DW_problem_3} and the linear interaction between axion and standard model operator~\cite{PhysRevLett.132.081003}.
In contrast to these model-building approaches, we propose a novel dynamical pathway in which primordial magnetic fields affect the evolution of the DW network in the early Universe. The primordial magnetic field (PMF), generated during inflation \cite{MF_inflation_1, MF_inflation_2, MF_inflation_3, MF_inflation_4, Ashoorioon_2005} and phase transitions ~\cite{MF_QCD,PhysRevLett.51.1488, Di_2021, Enqvist:1993np,Yang_2022,Vachaspati:1991nm,Baym:1995fk,Grasso:1997nx}, is widely considered as the seed of the observed cosmic magnetic field in  galaxy clusters~\cite{Clarke_2001, Bonafede_2010, Feretti_2012} and the voids of large-scale structures~\cite{Dolag_2010, Tavecchio_2010, Tavecchio_2011, Vovk_2012}.

In this Letter, we propose a dynamical mechanism in which the Chern--Simons (CS) coupling between the axion and a helical PMF provides an additional, spatially inhomogeneous driving force for the axion field. The resulting spatially nonuniform response changes the relative volumes of neighboring domains and thereby triggers the decay of the DW network. We perform large-scale three-dimensional lattice simulations of axion DWs coupled to a $U(1)$ gauge field, and, for the first time, we investigate in detail how the correlation length of a helical PMF affects the decay rate of DW networks, as well as the resulting axion dark matter and gravitational wave (GW) emission. We derive the resulting axion relic abundance and the corresponding constraints on the axion decay constant $f_a$ and the PMF correlation length $\lambda_B$. At the same time, we obtain a stochastic GW signal that falls in the range relevant for recent pulsar timing array experiments, such as NANOGrav \cite{gw_NANOGrav_1}, PPTA \cite{gw_PPTA_1}, and EPTA \cite{gw_EPTA_1}.

\noindent{\it \bfseries The simulation setup.}
We consider an axion model that couples to the gauge field through the CS term $F_{\mu\nu}\tilde{F}^{\mu\nu}$. Using conformal time $\tau$ and the FLRW metric $ds^2 = a^2
(\tau)(-d\tau^2 + dx^2)$, the action is
\begin{align}
\label{eq:action_axion}
S=\int d^4x\sqrt{-g}\Big[&-\frac{1}{2}\partial_{\mu}\phi\partial^{\mu}\phi-\frac{1}{4}F_{\mu\nu}F^{\mu\nu} \notag \\
&-V(\phi)-\alpha\frac{\phi}{4v} F_{\mu\nu}\tilde{F}^{\mu\nu}\Big],
\end{align}
where $\phi$ is the axion field and $\alpha$ is a dimensionless coupling constant. The parameter $v$ denotes the PQ vacuum expectation value, and the ratio $\theta=\frac{\phi}{v}$ can be viewed as the phase of the PQ field. The field strength tensor and its dual tensor are defined by $F_{\mu\nu}=\partial_{\mu}A_{\nu}-\partial_{\nu}A_{\mu}$, $\tilde{F}_{\mu\nu}=\frac{1}{2\sqrt{-g}}\epsilon_{\mu\nu\rho\sigma}F^{\rho\sigma}$, where $\epsilon_{\mu\nu\rho\sigma}$ is the Levi-Civita symbol and $\epsilon_{0123}=1$.
We use the potential
\begin{equation}
\label{eq:potential_axion}
V(\phi)=\frac{m_a(T)^2v^2}{N_{\rm DW}^2}\left(1-\text{cos}\left(N_{\rm DW}\frac{\phi}{v}\right)\right).
\end{equation}
For temperatures $T > 100\mathrm{MeV}$, the axion mass $m_a(T)$ can be parametrized as \cite{Wantz_2010}
\begin{equation}
\label{eq:axion_mass}
m_a(T)^2 = \frac{\alpha_a \Lambda^4}{f_a^2(T/\Lambda)^{6.68}},
\end{equation}
where $\alpha_a=1.68\times10^{-7}$, $\Lambda=400~\text{MeV}$ and $f_a=v/N_{\rm DW}$. When $T \leq 100 ~\text{MeV}$, the axion mass no longer grows and is replaced by a zero-temperature mass $m_{a,0} = 5.707 \times 10^{-5} (10^{11} \text{GeV}/ f_a ) \text{eV}$ \cite{di_Cortona_2016}.

This potential has $N_{\rm DW}$ degenerate minima, corresponding to $N_{\rm DW}$ distinct vacua. During the phase transition, the axion field settles into different vacua in different regions, and the boundaries between these regions after the transition form DWs. The width and tension of these DWs are $r_{\rm DW}\sim m_a^{-1}, \sigma_{\rm DW}\sim m_af_a^2$ \cite{texbook_cosmic_string}. The oscillation of DWs can be the source of GW, and the energy density of GW can be calculated by $\rho_{gw}\sim G\sigma_{dw}^2$ \cite{2406.17053}. The GW spectrum is computed within the lattice simulation using the Pystella codes introduced in Ref.~\cite{pystella}, with the relevant equations provided in the Supplemental Material. The DW network tends to evolve toward a scaling regime \cite{texbook_cosmic_string}, when the DW area per Hubble patch ($\mathcal{A}$) is a constant. However, deviations have been reported in some studies \cite{1310.1774, 1412.0789}. With the total area of DWs ($A$) and the total volume $V$, we define $\mathcal{A}$ as the scaling parameter
$   \mathcal{A}=At/V$. In our code, we use the same method as \cite{2311.20211} to identify the location of DWs.

Adopting the temporal gauge $A_0=0$, the equations of motion (EOM) are \cite{1705.09629, pystella}
\begin{align}
\label{eq:EOMs}
\phi^{\prime\prime}&=\partial_i\partial_i\phi-2\mathcal{H}\phi^{\prime}-a^2\frac{dV}{d\phi}-a^2\frac{\alpha}{4v}F_{\mu\nu}\tilde{F}^{\mu\nu}\notag ,\\
E_i^{\prime}&=-\epsilon_{ijk}\partial_jB_k+\frac{\alpha}{v}\phi^{\prime}B_{i}-\frac{\alpha}{v}\epsilon_{ijk}\partial_j\phi E_k \notag ,\\
\partial_iA^{\prime}_i&=\frac{\alpha}{v}\partial_i\phi\epsilon_{ijk}\partial_jA_k,
\end{align}
where $\phi^{\prime}=\partial\phi/\partial \tau$ and $A_i$ is the comoving gauge field. In addition, $E_i=A_i^{\prime}$ and $B_i=\epsilon_{ijk}\partial_jA_k$ are the comoving electric and magnetic fields and $\mathcal{H}=aH$ is the conformal Hubble. The third equation is the Gauss constraint, arising naturally from gauge invariance. The CS term is $F_{\mu\nu}\tilde F^{\mu\nu}=4a^{-4}\sum_iE_iB_i$, with the factor $a^{-4}$ accounting for cosmological dilution, which renders the source dynamically negligible at late times. The subsequent axion evolution is therefore governed by the usual periodic QCD potential, as discussed in the Supplemental Material. We use the discretization scheme mentioned in \cite{1705.09629} to evolve the above equations on the lattice.

In this work, we consider the PMF with finite correlation length $\lambda_B$ and helicity $\mathscr{H}$, which are defined by
\begin{equation}
\label{eq:lambda_H}
\lambda_B=\frac{\int dk E_B(k)/k}{\int dk E_B(k)}, \quad \mathscr{H}=\int dx^3 \boldsymbol{A}\cdot \boldsymbol{B},
\end{equation}
where $\int dk E_B(k)=\frac{1}{2V}\int d^3x|\vec{B}(x)|^2$ is the magnetic field energy density. The time derivative of helicity determines the spatially averaged CS term \cite{helicity_CS}
\begin{equation}
\label{eq:helicity_CS_text}
\frac{d\mathscr{H}}{d\tau}=2\int dx^3 \boldsymbol{E}\cdot \boldsymbol{B},
\end{equation}
The PMF has nonzero initial helicity. The subsequent helicity oscillations arise dynamically from the early-time evolution of the individual gauge field modes, thereby producing an oscillating CS source through Eq.~\ref{eq:helicity_CS_text} (see Supplemental Material). It is useful to characterize the CS contribution to the axion equation of motion by introducing the local driving source

\begin{equation}
    \mathcal{S}_{\rm CS}(\mathbf{x},\tau)
    \equiv
    -\frac{\alpha a^2}{4v}
    F_{\mu\nu}\widetilde{F}^{\mu\nu}
    =
    \overline{\mathcal{S}}_{\rm CS}(\tau)
    +
    \delta\mathcal{S}_{\rm CS}(\mathbf{x},\tau),
    \label{eq:CS_source}
\end{equation}
where $\left\langle\delta\mathcal{S}_{\rm CS}\right\rangle=0$. Since $dV/d\phi$ is periodic and the gauge-field equations depend on $\phi$ only through its derivatives, the coupled equations of motion remain invariant under the global discrete shift $\phi\rightarrow\phi+2\pi v/N_{\rm DW}$. This symmetry relates globally shifted solutions; it does not require the wall geometry or the relative volumes of neighboring domains to remain unchanged during evolution. A spatially inhomogeneous CS source can therefore drive different local responses and relative wall motion while preserving the degeneracy of the periodic axion potential. Eq.~\ref{eq:helicity_CS_text} shows that a time-dependent magnetic helicity produces a nonzero spatially averaged source $\overline{S}_{\rm CS}$. Its time integral induces a coherent displacement of the spatially averaged axion field. Meanwhile, the local CS source remains strongly inhomogeneous, so that $\delta S_{\rm CS}$ excites nonzero-momentum axion modes and generates different local responses across the DW network. The coherent displacement determines the background about which the axion field evolves, whereas the inhomogeneous response changes the relative volumes and geometries of neighboring domains. In the thin-wall description, after the common bulk response has been separated, the inhomogeneous component acts as a position-dependent force that drives the relative motion of different wall segments. Since the physical CS source redshifts as $a^{-4}$, these responses are generated mainly at early times, around DW formation. The averaged axion response and a thin-wall description are analyzed in detail in the Supplemental Material.

We simulate the evolution during the QCD phase transition. The initial temperature is set to $T_i = 220~\text{MeV}$, and the PQ vacuum is taken as $v = N_{\rm DW}f_a$, with $N_{\rm DW} = 3$ and $f_a = 1.8\times 10^{16} \text{GeV}$. The effective degrees of freedom $g(T)$, scale factor $a(T)$ and Hubble parameter $H(T)$ are determined using the expressions provided in \cite{2311.20211}. In the simulation, we set $f_* = v$ and $w_* = 6a_iH_i$, where $a_i$ and $H_i$ denote the initial values of $a(T)$ and $H(T)$, to define the following dimensionless variables
\begin{equation}
\label{eq:dimansionless_variety}
\tilde{\phi}=\frac{\phi}{f_*}, \quad \tilde{A_i}=\frac{A_i}{w_*}, \qquad\tilde{x}=w_*x, \quad \tilde{\tau}=w_*\tau.
\end{equation}
We use a lattice of size $N^3 = 1024^3$, corresponding to an initial volume containing roughly $10^3$ Hubble volumes and a final volume still exceeding one Hubble volume. The dimensionless spatial and temporal steps are set to $d\tilde{x} = 0.06$ and $d\tau = 0.012$, which ensures that the physical lattice spacing $dx$ remains smaller than the DW width $r_{dw}$ throughout the evolution. We evolve the simulation until the temperature reaches $T \approx 30\,\mathrm{MeV}$. For each parameter set, we perform five independent simulations with different initial random number seed. Numerical convergence tests demonstrate that our main results are robust against both resolution and finite-volume effects.

For the PMF with finite correlation length, it is defined in the momentum space \cite{PhysRevD.102.023536}:
$ B_i(\vec{k})=B_{ini}\Theta(k_{UV}-k)(\delta_{ij}-\hat{k}_i\hat{k}_j-i\sigma_M\epsilon_{ijl}\hat{k}_l)g_j(\vec{k})k^n$
where $k_{UV}$ is the cutoff for the PMF and $g_j(\vec{k})$ is the Fourier transform of a Gaussian distributed random vector field that is $\delta$-correlated in all three dimensions. The exponent $n$ is the spectral index and $\sigma_M\in[-1, 1]$ determines the helicity:
\begin{equation}
\label{eq:f_H_sigma_text}
f_H=\frac{\sum_kh(k)}{\sum_k\frac{2\rho_B(k)}{k}}=\frac{2\sigma_M}{1+\sigma_M^2},
\end{equation}
where $h(k)$ and $\rho_B(k)$ denote the helicity and energy density of the magnetic field in momentum space, respectively. This relation indicates that for a fixed magnetic energy density, the helicity is determined solely by the parameter $\sigma_M$. The detailed derivations and a comprehensive analysis of the initial axion and gauge fields are provided in the Supplemental Material. The spectral index of the initial magnetic field is fixed at $n=2$. 

We simulate the case without the CS term by setting $\alpha = 0$, and use $ \alpha = 0.2$ to study the PMF effects on DW networks when the CS term is present. We choose $B_{ini}$ such that the initial magnetic-field energy density is nearly the same across cases. The effects of $f_H$, $\alpha$, and $B_{ini}$ are discussed below.

\noindent{\it \bfseries Numerical results.} We find that the DW decay rate depends on both the temporal and spatial structure of the CS source, as shown in the Supplemental Material. For short correlation lengths, rapid oscillations of $\overline{\mathcal{S}}_{\rm CS}$ lead to a partial cancellation of the accumulated bulk impulse. For long correlation lengths, the averaged source is smaller because the magnetic helicity evolves more slowly. The dynamical response is therefore strongest at intermediate correlation lengths, where both the amplitude and the time scale of the source are favorable. Our Hubble-patch analysis further shows that the CS source contains a coherent coarse-grained component together with sizable local fluctuations. The former generates coherent motion of the spatially averaged axion field, whereas the latter drives differential evolution across the DW network and reshapes the walls.

We parameterize the decrease of the scaling parameter $\mathcal{A}$ as 
\begin{equation}
\label{eq:scaling_A_evolve}
\mathcal{A}(t)=\mathcal{A}_{\text{form}}\left(\frac{t}{t_{\text{form}}}\right)^{1-p},
\end{equation}
where we take the DW network to form at time $t_{\text{form}}$ with $\mathcal{A}=\mathcal{A}_{\text{form}}$. For $\alpha=0$, the parameter $p$ is approximately constant and slightly smaller than unity, reflecting a deviation from exact scaling \cite{1412.0789}. For $\alpha\neq0$, $p$ increases approximately linearly with conformal time, as shown in the Supplemental Material. We fit $p(t)$ linearly and use it to extrapolate the long-time evolution of $\mathcal{A}$ and determine the decay time, defined by $\mathcal{A} = 0.01\mathcal{A}_{\text{form}}$ or $\mathcal{A} = 0.1\mathcal{A}_{\text{form}}$. These decay criteria are widely adopted in the literature~\cite{1412.0789, PhysRevD.85.105020, JCAP08(2011)030}. The corresponding decay times for different $\lambda_B$ cases are shown in the Supplemental Material. Both criteria indicate that the relationship between the decay time $\log_{10}(\frac{t_{\text{dec}}}{t_{\text{form}}})$ and the correlation length $\log_{10}(\lambda_B H_i)$ is piecewise linear, from which we can also derive the corresponding fitting function. This piecewise linear behavior reflects the presence of different dominant factors in the long- and short-correlation regimes as discussed above.
For all considered $\sigma_M$, the parameter $p$ still exhibits a linear increase with time, allowing us to calculate the decay time as before.
By estimating $\log_{10}(\frac{t_{\text{dec}}}{t_{\text{form}}})$ for different values of $\sigma_M$, we find $\log_{10}(\frac{t_{\text{dec}}}{t_{\text{form}}}) \propto 1/f_H = \frac{1+\sigma_M^2}{2\sigma_M}$, showing that the decay time depends sensitively on the helicity.

The DW network radiates energy into free axions. In the pure axion case, the network nearly reaches a scaling regime, so most of the energy cannot be emitted as free axion, and the axion energy is related to the parameter $p$ \cite{1412.0789}. However, when gauge coupling is taken into account, the DWs decay rapidly, and we follow the standard approximation that the radiated DW energy is converted predominantly into free axions. Following the derivation in the Supplemental Material and using the same normalization parameters as in Ref.\cite{2212.13263}, we obtain the present-day axion relic abundance for the gauge-coupling cases: 

\begin{align}
\Omega_{a,0}
\sim & 0.0225 \frac{\kappa}{\epsilon_a}\mathcal{A}_{\text{form}} \left(\frac{t_{\text{dec}}}{t_{\text{form}}}\right)^{3/2}\left(\frac{10.75}{g_*(T_{\text{dec}})}\right)^{1/4}
\notag \\
&\times\left(\frac{f_a}{10^{12}\text{GeV}}\right)^2 \left(\frac{m_a}{ 10^{-6}\text{eV}}\right)^{1/2}.
\label{eq:ratio_axion_DM_text}
\end{align}

For each case, this expression requires the average momentum parameter 
\begin{equation}
    \epsilon_a = \frac{\omega_a}{m_a}, \quad \omega_a=\sqrt{\bar{k}^2+m_a^2}, \quad \bar{k} = \frac{\int \rho_a(k)dk}{\int (\rho_a(k)/k)dk}
\end{equation}
where $\bar{k}$ is the average momentum and $\mathcal{A}_{\text{form}}$ is the scaling parameter at formation. The dimensionless factor $\kappa$ parameterizes the influence of continuous DW emission and the time dependence of $p(t)$. These results are presented in the Supplemental Material. We fit $\epsilon_a$, $\mathcal{A}_{\rm form}$, and $t_{\rm dec}/t_{\rm form}$ as functions of the correlation length $\lambda_B$. The parameters entering Eq.~\ref{eq:ratio_axion_DM_text} are fitted as
\begin{align}
&\epsilon_a=0.05\left(\frac{C}{C_0}-1\right)+ \begin{cases}
0.05x+1.15,&  x\in R_1 \\
-0.11x+1.04,&  x\in R_2
\end{cases} \notag\\
&\mathcal{A}_{\text{form}}=0.60\left(\frac{C_0}{C}-1\right) +
\begin{cases}
-0.91x+1.72,&  x\in R_1 \\
0.29x+2.55,&  x\in R_2
\end{cases} \notag\\
&\log_{10}\left(\frac{t_{\text{dec}}}{t_{\text{form}}}\right)=0.83\left(\frac{C_0}{C}-1\right) +
\begin{cases}
-1.70x-0.29, x\in R_1 \\
1.28x+1.80, x\in R_2
\end{cases} \notag\\
&\log_{10}(\kappa)=1.26\left(\frac{C}{C_0}-1\right) +
\begin{cases}
1.42x+0.16,&  x\in R_1 \\
-1.06x-1.58,&  x\in R_2
\end{cases}
\label{eq:ratio_axion_DM_para}
\end{align}
with $x=\log_{10}(\lambda_BH_i)$, and we define the decay time as the moment when $\mathcal{A}=0.01\mathcal{A}_
{\text{form}}$. Here, $R_1$ and $R_2$ denote two ranges of $x$, with $R_1$ corresponding to $x<-0.7$ and $R_2$ to $x\geq -0.7$. For cases with fixed $\sigma_M=1$, both $\epsilon_a$ and $\mathcal{A}_{\rm form}$ exhibit approximately piecewise-linear dependence on $\log_{10}(\lambda_B H_i)$. This behavior reflects the correlation-length dependence of the accumulated CS impulse. A stronger axion response, together with the spatial variation of the CS source, enhances the local reconfiguration of the field, thereby reducing the total DW area. Additionally, we find $\mathcal{A}_{\text{form}} \propto 1/f_H$ and $\epsilon_a \propto f_H$ for different $\sigma_M$ with fixed $\lambda_B \sim 0.2 H_i^{-1}$ in the Supplemental Material. 

The parameter $C$ in Eq.~\ref{eq:ratio_axion_DM_para} measures the coupling strength. For the fixed energy density of the PMF, the CS term satisfies $F_{\mu\nu}\tilde{F}^{\mu\nu} \propto f_H$. Because the gauge field is initialized as $A_i \propto B_{ini}$, the CS term scales accordingly, yielding $F_{\mu\nu}\tilde{F}^{\mu\nu} \propto B_{ini}^2$. In addition, when the CS term is fixed, the effect of the coupling constant $\alpha$ appears solely through the coupling term, meaning that the influence of $\alpha$ follows the same scaling behavior as $f_H$. Therefore, we parameterize the coupling term $C$ as:
\begin{equation}
\label{eq:para_coupling}
C=C(\alpha, B_{\text{ini}}, f_H)=\left(\frac{\alpha}{\alpha_0}\right)\left(\frac{B_{ini}^2}{B_0^2}\right)\left(\frac{f_H}{f_{H,0}}\right) C_0,
\end{equation}
where $\alpha_0=0.2, f_{H,0}=1$ are the reference values adopted in our simulations, and $C_0$ corresponds to $(\alpha,B_{\rm ini},f_H)=(\alpha_0,B_0,f_{H,0})$. We provide a conservative estimate of the free-streaming scale in the Supplemental Material and find that the emitted axions remain safely consistent with cold-dark-matter behavior. To avoid DW domination, we require $t_{\rm dec}<t_{\rm WD}$, where $t_{\rm WD}(f_a)=3/(16\pi G\sigma_{\rm DW}(f_a))$ denotes the time at which DWs would otherwise dominate the Universe \cite{2311.20211}. With $t_{\rm dec}$ given by Eq.~\eqref{eq:ratio_axion_DM_para}, we have
\begin{equation}
    \frac{C(\alpha,B_{\rm ini}, f_H)}{C_0}>\frac{0.83}{L(f_a)+0.83-F(x)},
\end{equation}
provided that the denominator is positive. Here, $F(x)$ is the $x$-dependent part of the fit for $\log_{10}(t_{\rm dec}/t_{\rm form})$, and $L(f_a)=\log_{10}[t_{\rm WD}(f_a)/t_{\rm form}]$.

\begin{figure} [!htp]
\includegraphics[width=0.45\textwidth]{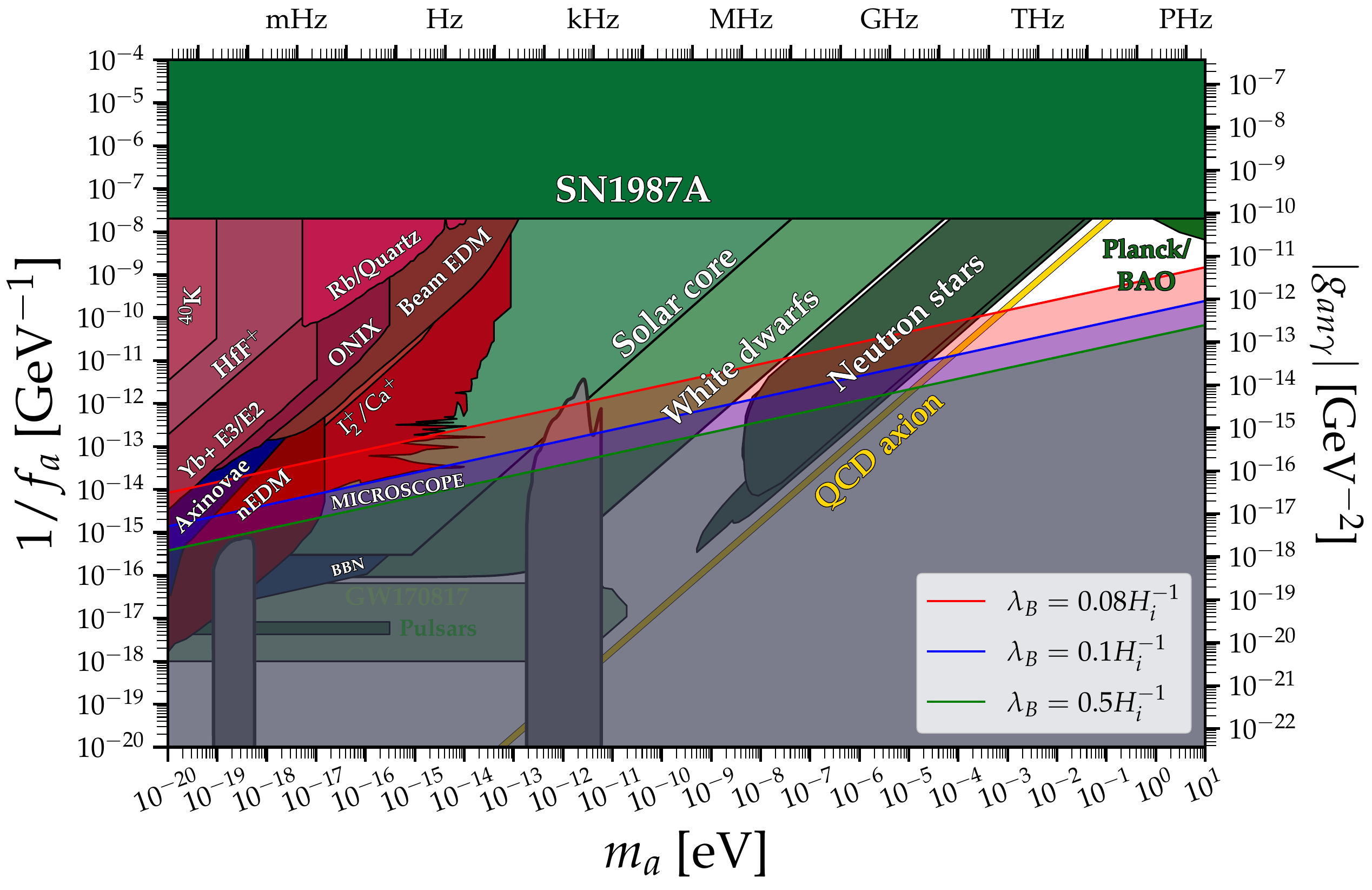}
\caption{The parameter space of $f_a$ and $m_a$. The shaded region represents the forbidden region $\Omega_{a}>\Omega_{\rm DM}$. The data used here are from \cite{AxionLimits}. We take $C/C_0\approx 0.27, 0.42$ and $1$ for $\lambda_B=0.08H_i^{-1}, 0.1H_i^{-1}$ and $0.5H_i^{-1}$ cases, respectively.}
\label{fig:axion_constraint(10.30)}
\end{figure}

Using the constraint $\Omega_{a,0} / \Omega_{\text{DM}} \leq 1$, we obtain the allowed parameter space of $f_a$ and $m_a$ for different correlation lengths $\lambda_B$, which is presented in Fig.~\ref{fig:axion_constraint(10.30)}. Since the PMF amplitude used in our simulations exceeds observational bounds for some $\lambda_B$ by an $O(1)$ factor, we rescale $B_{ini}$ (and hence $C$) to saturate the bounds. The colored areas denote regions of the axion parameter space that are excluded by existing laboratory, astrophysical, and cosmological constraints, and the shaded region represents the forbidden region $\Omega_{a,0}>\Omega_{DM}$. In addition, the bound $t_{\rm dec}<t_{\rm WD}$ is significantly weaker than the constraint derived from requiring $\Omega_{a,0} / \Omega_{\text{DM}} \leq 1$. As shown in Fig.~\ref{fig:axion_constraint(10.30)}, shorter correlation lengths yield stricter constraints. A smaller correlation length slows the decay of the DW network, so axion production occurs later and undergoes less redshifting. This leads to a larger axion energy density today, thereby tightening the corresponding parameter constraints.

Next, we consider the GW emission of the DW network. Near the left peak, the GW spectra follow power-law dependencies $k^{2.05}$ in the low-frequency regime and $k^{-1.05}$ in the mid-frequency regime. The peak value of the GW spectra grows approximately as $H^{-2}(\tau)$, consistent with the results of \cite{2406.17053}. Using $\rho_{gw}\sim G\sigma_{dw}^2\sim Gf_a^4m_a^2$, the peak value can be parameterized as $\Omega_{gw}^{p}(\tau)\propto H^{-2}(\tau)f_a^4m_a^2$. The GW amplitude also depends on the instantaneous wall area $\mathcal{A}$. Since the network is not in the scaling regime and $\mathcal{A}$ varies strongly with time, we absorb this effect into the fitted overall coefficient. As detailed in the Supplemental Material, the present-day peak amplitude is 
\begin{align}
\label{eq:Omega_text}
\Omega_{gw,0}^{p}=
\tilde{\Omega}  \left(\frac{t_{\text{dec}}}{t_{\text{form}}}\right)^2 \left(\frac{f_a}{1.8\times 10^{16}\text{GeV}}\right)^4 \left( \frac{10.75}{g_*(T_{\text{dec}})} \right)^{{4/3}},
\end{align}
where the factor $\tilde{\Omega}$ is given by the following piecewise-linear function:
\begin{align}
&\tilde{\Omega}= 10^{-14}\left(1.39\left(\frac{C_0}{C}-1\right)
+\begin{cases}
-2.82x+1.59, & x\in R_1 \\
0.85x+4.16, & x\in R_2
\end{cases}\nonumber
\right)
\end{align}
where $x=\log_{10}(\lambda_BH_i)$. We compute the GW efficiency parameter $\epsilon_{gw} = \rho_{gw}/(G\sigma_{dw}^2)$, finding that it is approximately an order of magnitude smaller than the value reported in Ref.~\cite{1207.3166}, but it is close to the results presented in Ref.\cite{2406.17053}. We also investigate the effect of $\sigma_M$ on GW emission for the case $\lambda_B \sim 0.2H_i^{-1}$, as presented in the Supplemental Material. The results show that $\Omega_{gw,0}^{p}h^2 \propto 1/f_H$, which is consistent with the expectation that a faster-decaying DW network produces weaker GW radiation.

\begin{figure}[!htp]
\centering
\begin{subfigure}{0.4\textwidth}
\includegraphics[width=\textwidth]{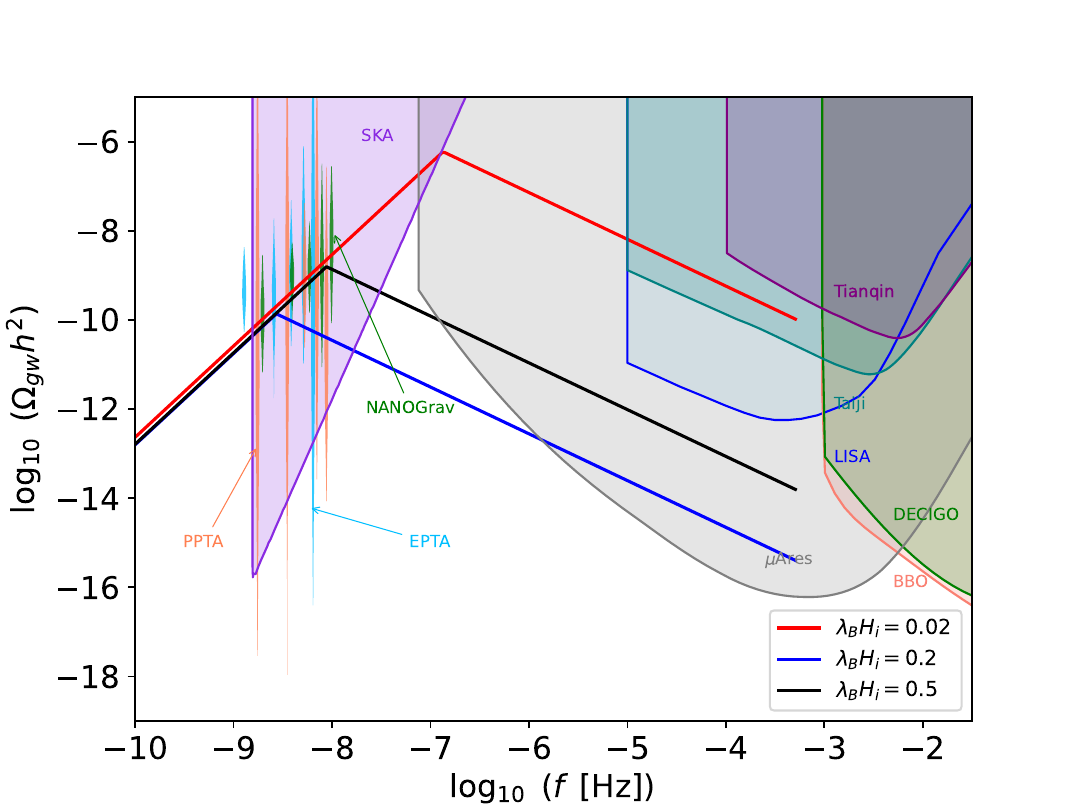}
\end{subfigure}
\caption{Predicted GW spectra for representative $\lambda_B$ cases with $f_a=5\times10^{16}~\text{GeV}$ corresponding to a zero-temperature axion mass of $m_{a,0}=1.14\times10^{-10}\,\mathrm{eV}$. The sensitivity curves of LISA~\cite{gw_LISA_1, gw_LISA_2}, TianQin \cite{gw_TianQin_1, gw_TianQin_2}, Taiji \cite{gw_TaiJi_1, gw_TaiJi_2}, $\mu$Ares \cite{gw_muAres_1}, DECIGO \cite{gw_DECIGO_1,gw_DECIGO_2,gw_DECIGO_3,gw_DECIGO_4}, BBO \cite{gw_BBO_1,gw_BBO_2,gw_BBO_3}, SKA \cite{gw_SKA_1}, NANOGrav \cite{gw_NANOGrav_1}, PPTA \cite{gw_PPTA_1} and EPTA \cite{gw_EPTA_1} are presented.}
\label{fig:gw_constraint(10.30)}
\end{figure}

Combining the peak amplitude with the spectral slopes on both sides, we obtain the present-day GW spectra shown in Fig.~\ref{fig:gw_constraint(10.30)} for $f_a=5\times 10^{16}$ GeV. The predicted spectra can overlap the sensitivity ranges of the stochastic background reported by PTA experiments (e.g., NANOGrav, PPTA, EPTA, and CPTA), although a dedicated fit to the PTA spectral shape is beyond the scope of the present work; see Refs.~\cite{Bian:2020urb,Bian:2023dnv} for related analyses based on the spectra of Ref.~\cite{1207.3166}. However, this benchmark is not viable in the minimal QCD axion scenario due to dark matter overproduction. Imposing the dark-matter constraint suppresses the corresponding GW amplitude well below current PTA sensitivity.

\noindent{\it \bfseries Conclusion and Discussion.} Using three-dimensional lattice simulations of axion DW networks interacting with a primordial magnetic field through the CS term during the QCD epoch, we propose a novel resolution to the domain wall problem. This solution relies on oscillations of the magnetic helicity, which generate a transient and spatially structured CS source. Its spatially averaged component produces a bulk displacement of the axion field, while its inhomogeneous component produces different local responses across the DW network. The combined evolution changes the relative volumes of the domains, deforms their boundaries, and eventually drives the network to collapse, while the degenerate vacuum structure of the axion potential remains unchanged.

We find that the decay rate of the DW network depends sensitively on the PMF correlation length $\lambda_B$ and the helicity strength. Short-correlation PMFs lead to near cancellation of the axion velocity due to rapid oscillations of the Chern-Simons term, whereas long-correlation PMFs result in only a small net Chern-Simons contribution because of slow helicity oscillations; both regimes delay DW decay. Based on the fitted $\lambda_B$-dependence of parameters such as the decay time, we quantify axion particle production and derive dark matter constraints on $f_a$ and $m_a$ for each $\lambda_B$. We further examine the influence of the coupling constant $\alpha$, the helicity fraction $f_H$, and the initial magnetic-field strength $B_{ini}$ on dark matter production. Following Ref.~\cite{2311.20211}, a larger $N_{\rm DW}$ may mildly enhance $\mathcal{A}_{\rm form}$, while the decay time is expected to be mainly controlled by the gauge coupling. We therefore expect only an $O(1)$ effect on the axion abundance, leaving a detailed study of the $N_{\rm DW}$ dependence to future work. Previous analytical studies considered axion production from a spatially uniform axion field coupled to a helical PMF~\cite{PhysRevLett.96.161302} and showed that a coherent helical electromagnetic background can induce a net axion drift through the CS coupling~\cite{Kobayashi_2021}. In this work, we go beyond the homogeneous approximation by employing high-resolution numerical simulations that track the evolution of the axion field from a more realistic initial configuration. Additionally, we present GW spectra for different $\lambda_B$ and derive the corresponding GW radiation, providing correlated predictions for axion DM and GW signals that may be tested by future observations, including pulsar timing array experiments.

After accounting for cosmic expansion and interactions with the turbulent plasma~\cite{Durrer_2013}, our simulation parameters correspond to present-day values of approximately $\lambda_B\sim 0.01$--$0.1\,{\rm Mpc}$ and $|B|\sim 5.7\times10^{-10}\,{\rm G}$. In our simulations, the gauge coupling enhances the magnetic field amplitude, with a stronger effect for longer correlation lengths. For instance, $|B|$ is about $1.3$ times larger than that in the no-coupling case for $\lambda_B \sim 0.5H_i^{-1}$. We further note that the pseudo-scalar dark matter, coupled to a gauge field via the CS term, could generate helical magnetic fields following recombination~\cite{2502.19310}. It was also found that axion–photon coupling may induce a rotational component in thermoelectric fields during the QCD crossover, thereby seeding primordial magnetic fields~\cite{Miniati_2018}. These mechanisms provide promising avenues for cosmological magnetic-field generation, which we leave for future numerical studies. 

We finally clarify a limitation of our simulations: the gauge-field equations do not include an Ohmic current and therefore describe the zero-conductivity limit. If the simulated $U(1)$ field is identified with the Standard Model electromagnetic field at the QCD epoch, the high conductivity of the primordial plasma would damp electric fields and thereby reduce the CS source, $S_{\rm CS}\propto \boldsymbol{E}\cdot\boldsymbol{B}$ \cite{high-conductivity}. The DW-decay efficiency obtained here may consequently overestimate that in the visible sector, and our results should be interpreted as a zero-conductivity benchmark for this case.
Alternatively, the simulated field may represent a dark $U(1)$ gauge field coupled to the axion through the same CS interaction \cite{0810.5126, 2511.15810}. In this interpretation, the conductivity depends on the dark charged-particle content. The present approximation is self-consistent if the dark sector contains no abundant light charged particles capable of sustaining appreciable currents. Future quantitative studies may incorporate finite conductivity and its backreaction on axion–gauge-field dynamics, building on the benchmark presented here.

\noindent{\it \bfseries Acknowledgments.}
This work is supported by the National Key Research and Development Program of China under Grant No.2021YFC2203004, and by the National Natural Science Foundation of China (NSFC) under Grants Nos.12322505, 12547101. We also acknowledge Chongqing Natural Science Foundation under Grant No. CSTB2024NSCQJQX0022 and Chongqing Talents: Exceptional Young Talents Project No. cstc2024ycjhbgzxm0020. J.S. is supported by Peking University under Startup Grant No.7101302974, the NSFC under Grants No.12025507 and No.12450006.

\bibliographystyle{apsrev4-1}
\bibliography{references}

\clearpage

\onecolumngrid
\begin{center}
  \textbf{\large Supplemental Material}\\[.2cm]
\end{center}

\noindent{\it \bfseries Discrete EOMs} To numerically solve Eq.4 in the main text on the lattice, we adopt the standard staggered grid and leap-frog discretization. The scalar field and the gauge field are defined on alternating lattice sites, $(n_0,\boldsymbol{n})$ and $(n_0,\boldsymbol{n}+\frac{\boldsymbol{i}}{2})$, respectively.

\begin{align}
\label{eq:physical_sites}
&A_i(x,t)\rightarrow(n_0,\mathbf{n}+\frac{\mathbf{i}}{2}), \quad \phi(x,t)\rightarrow(n_0+\frac{1}{2},\mathbf{n}),\quad \phi^{\prime}(x,t)=\partial_0^-\phi\rightarrow(n_0,\mathbf{n}),\notag \\
&B_i(x,t)=\nabla^+\times \mathbf{A}\rightarrow(n_0,\mathbf{n}+\frac{\mathbf{j}}{2}+\frac{\mathbf{k}}{2}),\quad E_i(x,t)=\partial_0^+A_i\rightarrow(n_0+\frac{1}{2},\mathbf{n}+\frac{\mathbf{i}}{2}),
\end{align}
where the finite-difference operators $\partial_0^\pm$ and $\nabla^\pm$ correspond to forward and backward differences, respectively. This definition in the time direction is the same as the leapfrog method.

To ensure that each update involves only quantities located at the same lattice site, we introduce suitable local averaging and interpolation schemes, leading to the discrete evolution equations for the scalar and gauge momenta given in
\begin{align}
    \label{eq:EOM_scalar_dis}
    \phi^{\prime}(\tau+\Delta\tau)=\phi^{\prime}(\tau)\frac{a^2(\tau)}{a^2(\tau+\Delta\tau)}+\frac{d\tau}{a^2(\tau+\Delta\tau)}&\Big [a^2(\tau)\nabla^+\nabla^-\phi(\tau)- a^4(\tau)\frac{\partial V(\phi)}{\partial\phi}(\tau) \notag \\
    &-\frac{\alpha}{v}a^4(\tau)E_i^{(2)}(\tau)(B_i^{(4)}(\tau)+B_i^{(4)}(\tau+\Delta\tau))\Big],
\end{align}

\begin{align}
    \label{eq:EOM_gauge_dis}
    A_i^{\prime}(\tau)=A_i^{\prime}(\tau-\Delta\tau)+d\tau &\Big[-\epsilon_{ijk}\partial^-_jB_k(\tau)+\frac{\alpha}{2v}\Big(\phi^{\prime}(\tau)B_i(\tau)+\phi^{\prime}_{+i}(\tau)B_{i,+i}(\tau)\Big) -\frac{\alpha}{4v}\epsilon_{ijk}\Big(K_{jk}+K_{jk,+i}\Big)\Big],
\end{align}
where we define
\begin{align}
\label{eq:physical_sites_2}
&B^{(4)}_i(x)=\frac{1}{4}[B_i(x)+B_{i,+j}(x)+B_{i,+k}(x)+B_{i,+j+k}(x)]\rightarrow(n_0,\mathbf{n}), \notag \\
&E^{(2)}_i(x)=\frac{1}{2}[E_i(x)+E_{i,+i}(x)]\rightarrow(n_0+\frac{1}{2},\mathbf{n}),
\notag \\
&K_{jk}=\partial^-_j\phi(\tau)E_{k,-j}+\partial^+_j\phi(\tau)E_{k,+j}(\tau),
\end{align}
Here $E_{i,+i}(x,\tau)=E_i(x+\hat{i},\tau)$ and $\phi_{+i}(x,\tau)=\phi(x+\hat{i},\tau)$ denote fields shifted by one lattice spacing in the $i$ direction. In addition, the discrete Gauss constraint is
\begin{equation}\label{eq:Gauss_dis}
\partial^-_iA^{\prime}_i(\tau+\Delta\tau)=\frac{\alpha}{2v}\Big[\partial^-_i\phi(\tau+\Delta\tau)\epsilon_{ijk}\partial^+_jA^{(2)}_{k,-i}(\tau)+\partial^+_i\phi(\tau+\Delta\tau)\epsilon_{ijk}\partial^+_jA^{(2)}_{k,+i}(\tau)\Big],
\end{equation}
where $A_i^{(2)}(x)=\frac{1}{2}[A_i(x)+A_i(x+i)]$. Finally, the fields $\phi, A_i$ are updated as
\begin{align}\label{eq:EOM_fields_dis}
&\phi(\tau+\Delta\tau)=\phi(\tau)+\Delta\tau \phi^{\prime}(\tau+\Delta\tau), \notag \\
&A_{i}(\tau+\Delta\tau)=A_i(\tau)+\Delta\tau A^{\prime}_i(\tau),
\end{align}

This construction provides a stable and constraint-preserving integration scheme suitable for long-term lattice simulations \cite{1705.09629}.

\noindent{\it \bfseries Initial condition} For the axion field, the dimensionless variable $\theta=\phi/v$ corresponds to the phase of the PQ field. Since its value is randomly selected during PQ symmetry breaking, we initialize $\theta$ with a uniform distribution in the range $[0,2\pi)$. A dimensionless momentum cutoff $\tilde{k}_{cut}=1$ is imposed so that the axion field correlation length is $\lambda_a\sim 0.1H^{-1}$. For the gauge field, we first use $ B_i(\vec{k})=B_{ini}\Theta(k_{UV}-k)(\delta_{ij}-\hat{k}_i\hat{k}_j-i\sigma_M\epsilon_{ijl}\hat{k}_l)g_j(\vec{k})k^n$ to assign the initial magnetic field and then obtain the corresponding gauge field $A_i(\boldsymbol{x})$ and its conjugate momentum $A_i^{\prime}(\boldsymbol{x})$ using the method described in Ref.~\cite{2409.16124}.

In the PMF definition, we can vary the UV cutoff $k_{\text{UV}}$ to change the magnetic field’s correlation length. It is thus necessary to identify physically meaningful values of correlation length relevant to the primordial magnetic field at the QCD phase transition. We consider two possible origins for the primordial magnetic field: the QCD phase transition and inflation. The QCD-sourced magnetic field typically has a correlation length $\lambda_B \sim 0.1 H^{-1}$ \cite{PhysRevD.87.083007, 2009.14174} and a spectral index $n = 2$ \cite{Durrer_2013}. For inflation-generated fields, the comoving correlation length depends on the production time and can be stretched to super-Hubble scales if the field is generated sufficiently early. In this case, there is no upper bound on the comoving correlation length. Conversely, fields generated near the end of inflation may have a correlation length that, even after subsequent plasma effects \cite{Durrer_2013}, remains far below our lattice spacing at the QCD epoch and is therefore unresolved. Therefore, inflation allows us to explore a wide range of correlation lengths, both above and below the Hubble scale, and study their effects on the field evolution. The spectral index in this case depends on the specific inflationary model \cite{Durrer_2013}. We therefore vary the correlation length while fixing the spectral index to $n=2$ to isolate its effects.

In addition to the correlation length, the strength of the magnetic field must also be specified. A maximally helical PMF undergoes an inverse cascade during its interaction with the turbulent plasma, such that its comoving correlation length and comoving field strength evolve approximately as $\lambda_B \propto \tau^{2/3}$ and $\tilde B_\lambda \propto \tau^{-1/3}$, respectively \cite{Durrer_2013}. Once MHD turbulence has decayed, the subsequent evolution is governed by flux freezing and cosmological expansion, leading to an upper bound of the form $\tilde B_\lambda \lesssim 10^{-8} (\lambda_B/1~\mathrm{Mpc})\mathrm{G}$ on the present-day field parameters. In addition, the cosmic microwave background (CMB) observations impose independent limits on the magnetic field. Both the MHD decay bound and the CMB constraints are summarized in Ref.~\cite{Durrer_2013}. In our simulations, we choose the magnetic energy density to be $\rho_B \simeq 0.06\rho_{rad}$ at the initial time. Evolving this configuration through the MHD inverse cascade and subsequent cosmological expansion yields a present-day field strength and correlation length that are consistent with these observational bounds up to a factor of $\mathcal{O}(1)$.

\noindent{\it \bfseries The helicity.} With the total helicity $\mathscr{H}=\int dx^3\boldsymbol{A}\cdot \boldsymbol{B}$, we obtain
\begin{align}
    \label{eq:helicity_CS}
    \frac{d\mathscr{H}}{d\tau}&=\int dx^3 \boldsymbol{E}\cdot \boldsymbol{B} + \int dx^3 \boldsymbol{A}\cdot \frac{d}{d\tau}(\nabla \times \boldsymbol{A}) \notag \\
    &=\int dx^3 \boldsymbol{E}\cdot \boldsymbol{B} + \int dx^3 \boldsymbol{E}\cdot (\nabla \times \boldsymbol{A})-\int dx^3 \nabla\cdot(\boldsymbol{A}\times \boldsymbol{E}) \notag \\
    &=2\int dx^3 \boldsymbol{E}\cdot \boldsymbol{B}
\end{align}
where we have used the vanished integral on the boundary. This shows that a time-varying helicity leads to a nonzero spatial average of the Chern-Simons term, $F_{\mu\nu}\tilde F^{\mu\nu}=\frac{4}{a^4}\mathbf E\cdot\mathbf B$.

For the initial magnetic field defined in $ B_i(\vec{k})=B_{ini}\Theta(k_{UV}-k)(\delta_{ij}-\hat{k}_i\hat{k}_j-i\sigma_M\epsilon_{ijl}\hat{k}_l)g_j(\vec{k})k^n$, the parameter $\sigma_M$ determines the magnitude of magnetic helicity. We derive this relationship in detail below. In momentum space, we introduce the transverse orthogonal basis vectors
\begin{equation}
\label{eq:momentum_base_vec}
\hat{k}\cdot\hat{e}_{\pm}=0, \quad \hat{e}_+\cdot \hat{e}_-=0, \quad i\hat{k}\times\hat{e}_{\pm}=\pm \hat{e}_{\pm},
\end{equation}
The magnetic field can then be decomposed as $\boldsymbol{B}(\boldsymbol{k})=B^+(\boldsymbol{k})\hat{e}_++B^-(\boldsymbol{k})\hat{e}_-$. The corresponding helicity and energy density in momentum space are given by
\begin{equation}
\label{eq:helicity_energy_B}
h(\boldsymbol{k})=\boldsymbol{A}(\boldsymbol{k})\cdot\boldsymbol{B}(\boldsymbol{k})^*=\frac{1}{k}(|B^+(\boldsymbol{k})|^2-|B^-(\boldsymbol{k})|^2), \quad \rho_B(\boldsymbol{k})=\frac{1}{2}\boldsymbol{B}(\boldsymbol{k})\cdot\boldsymbol{B}(\boldsymbol{k})^*=\frac{1}{2}(|B^+(\boldsymbol{k})|^2+|B^-(\boldsymbol{k})|^2),
\end{equation}
The total helicity fraction is therefore
\begin{equation}
\label{eq:ratio_f_H}
f_H=\frac{\sum_k h(k)}{\sum_k \frac{2\rho_B(k)}{k}}\in[-1, 1].
\end{equation}

We further define the transverse and antisymmetric projection operators as
\begin{equation}
\label{eq:operators}
P_{ij}^T=\delta_{ij}-\hat{k}_i\hat{k_j}, \quad P_{ij}^A=\epsilon_{ijl}\hat{k}_l.
\end{equation}
With these operators, the initial magnetic field can be expressed as $B_i(\boldsymbol{k})=B_{ini}\Theta(k_{UV}-k)(P_{ij}^T-i\sigma_MP_{ij}^A)g_j(\boldsymbol{k})k^n$. Noting that $P^{\pm} = \frac{1}{2}(P^T \mp iP^A)$ are the projection operators onto the $\hat{e}_{\pm}$ basis, the field components can be written as
\begin{equation}
\label{eq:B_plus_minus}
B^{\pm}_i(\boldsymbol{k})=B_{ini}\Theta(k_{UV}-k)(1\pm\sigma_M)P_{ij}^{\pm}g_j(\boldsymbol{k})=B_{ini}\Theta(k_{UV}-k)(1\pm\sigma_M)g_i^{\pm}(\boldsymbol{k}),
\end{equation}
where $g_i^{\pm}=P_{ij}^{\pm}g_j$ are the components of a Gaussian random field projected onto $\hat{e}_{\pm}$. Since $g_j$ is statistically isotropic and non-helical, we have $\sum_k |g_i^+(k)|^2 = \sum_k |g_i^-(k)|^2$. Substituting these relations into Eq.~\ref{eq:ratio_f_H} above, we obtain the final relation between $f_H$ and $\sigma_M$:
\begin{equation}
\label{eq:f_H_sigma}
f_H=\frac{\sum_k h(k)}{\sum_k \frac{2\rho_B(k)}{k}}=\frac{(1+\sigma_M)^2-(1-\sigma_M)^2}{(1+\sigma_M)^2+(1-\sigma_M)^2}=\frac{2\sigma_M}{1+\sigma_M^2},
\end{equation}

\noindent{\it \bfseries The origin of the CS source and its oscillation time scale.}
We now estimate the characteristic time scale and amplitude of the
oscillating CS source. Although the initial electric
field is not identically zero, our initialization prescription
introduces only the longitudinal component required by the Gauss
constraint. Since the magnetic field is transverse, this longitudinal
component satisfies
\begin{equation}
    \left\langle
    \boldsymbol{E}_{L}\cdot\boldsymbol{B}
    \right\rangle = 0 ,
    \label{eq:longitudinal_EB}
\end{equation}
and therefore does not contribute to the volume-averaged magnetic helicity evolution. The subsequent nonlinear effects of this longitudinal component are fully included in the lattice simulation; however, the component itself can be neglected in the following leading-order estimate of the oscillation timescale.

We decompose the transverse gauge field into circular-polarization
modes,
\begin{equation}
    \boldsymbol{A}(\boldsymbol{k},\tau)
    =
    \sum_{\lambda=\pm}
    A_{\lambda}(\boldsymbol{k},\tau)
    \boldsymbol{e}_{\lambda}(\boldsymbol{k}),
    \qquad
    i\boldsymbol{k}\times\boldsymbol{e}_{\lambda}
    =
    \lambda k\boldsymbol{e}_{\lambda}.
    \label{eq:helicity_decomposition}
\end{equation}

To identify the physical origin and characteristic time scale of the oscillating CS source, it is useful first to consider the free-field evolution of the transverse gauge-field modes. In this limit, each mode obeys
\begin{equation}
    A_{\lambda}''(\boldsymbol{k},\tau)
    +
    k^2 A_{\lambda}(\boldsymbol{k},\tau)
    \simeq 0 .
    \label{eq:free_gauge_mode}
\end{equation}
Because the initial transverse electric field vanishes under our initialization prescription, the leading-order solution is
\begin{equation}
    A_{\lambda}(\boldsymbol{k},\tau)
    \simeq
    A_{\lambda,i}(\boldsymbol{k})
    \cos\left[k(\tau-\tau_i)\right].
    \label{eq:gauge_mode_solution}
\end{equation}

The comoving magnetic helicity density is
\begin{equation}
    \mathscr{H}_B(\tau)
    \equiv
    \frac{1}{V}
    \int d^3x\,
    \boldsymbol{A}\cdot\boldsymbol{B}
    \label{eq:magnetic_helicity_density}
\end{equation}
Using Eq.~\eqref{eq:gauge_mode_solution}, it can be written in terms of
the initial helicity spectrum $h_i(k)$ as
\begin{equation}
    \mathscr{H}_B(\tau)
    =
    \int dk\,
    h_i(k)
    \cos^2\left[k(\tau-\tau_i)\right],
    \label{eq:helicity_evolution}
\end{equation}
where
\begin{equation}
    \mathscr{H}_B(\tau_i)
    =
    \int dk\,h_i(k).
\end{equation}

With our convention $\boldsymbol{E}=\boldsymbol{A}'$, periodic
boundary conditions give
\begin{equation}
    \frac{d\mathscr{H}_B}{d\tau}
    =
    2
    \left\langle
    \boldsymbol{E}\cdot\boldsymbol{B}
    \right\rangle .
    \label{eq:helicity_derivative}
\end{equation}
The physical CS term is therefore
\begin{equation}
    \left\langle
    F_{\mu\nu}\widetilde{F}^{\mu\nu}
    \right\rangle
    =
    \frac{4}{a^4}
    \left\langle
    \boldsymbol{E}\cdot\boldsymbol{B}
    \right\rangle
    =
    \frac{2}{a^4}
    \frac{d\mathscr{H}_B}{d\tau}.
    \label{eq:FFtilde_helicity}
\end{equation}
Substituting Eq.~\eqref{eq:helicity_evolution}, we obtain
\begin{equation}
    \left\langle
    F_{\mu\nu}\widetilde{F}^{\mu\nu}
    \right\rangle
    =
    -\frac{2}{a^4}
    \int dk\,
    k h_i(k)
    \sin\left[2k(\tau-\tau_i)\right].
    \label{eq:FFtilde_oscillation}
\end{equation}
Thus, the oscillating CS term is generated dynamically by the evolution of the initially helical magnetic configuration; it is not imposed as an independent initial condition. 

The helicity spectrum is related to the comoving magnetic-energy spectrum $\rho_{B,c}(k)$ by
\begin{equation}
    h_i(k)
    =
    \frac{2f_H(k)}{k}\rho_{B,c}(k),
    \label{eq:helicity_energy_relation}
\end{equation}
where $f_H(k)$ is the helicity fraction. Equation
\eqref{eq:FFtilde_oscillation} then becomes
\begin{equation}
    \left\langle
    F_{\mu\nu}\widetilde{F}^{\mu\nu}
    \right\rangle
    =
    -\frac{4}{a^4}
    \int dk\,
    f_H(k)\rho_{B,c}(k)
    \sin\left[2k(\tau-\tau_i)\right].
    \label{eq:FFtilde_energy_spectrum}
\end{equation}
For a maximally helical and approximately monochromatic magnetic
field, this gives the order-of-magnitude estimate
\begin{equation}
    \left|
    \left\langle
    F_{\mu\nu}\widetilde{F}^{\mu\nu}
    \right\rangle
    \right|_{\mathrm{peak}}
    \sim
    4\rho_B,
    \label{eq:FFtilde_peak}
\end{equation}
where
\begin{equation}
    \rho_B
    =
    \frac{1}{a^4}
    \int dk\,\rho_{B,c}(k)
\end{equation}
is the physical magnetic-field energy density. For a broad spectrum,
the actual peak is smaller than Eq.~\eqref{eq:FFtilde_peak} because
different Fourier modes dephase.

\begin{figure}[!htp]
\centering
\begin{subfigure}{0.4\textwidth} \includegraphics[width=\textwidth]{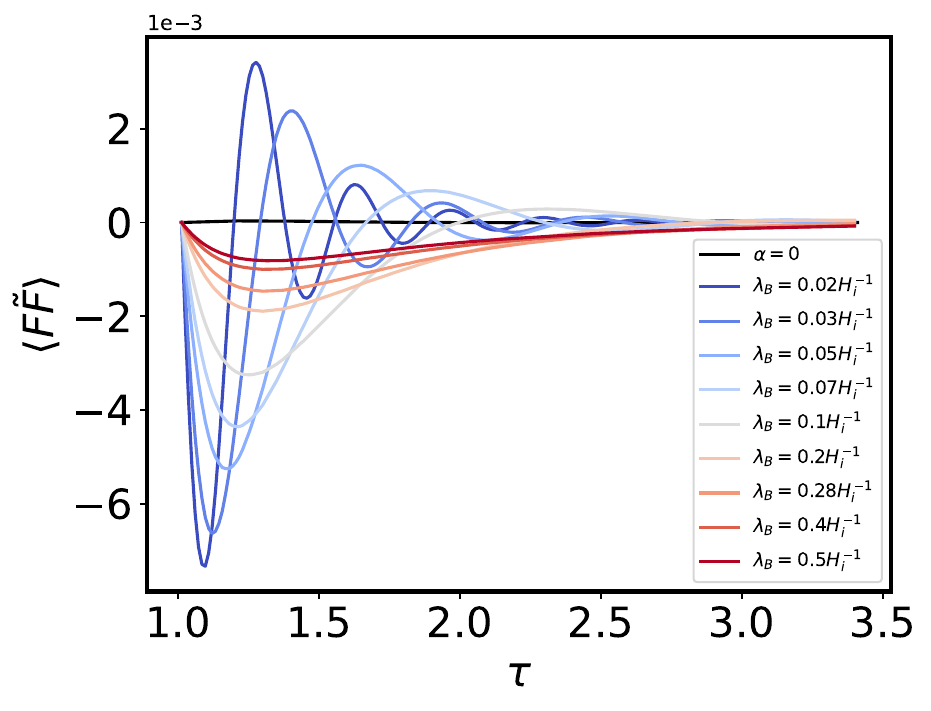}
\end{subfigure}
\begin{subfigure}{0.43\textwidth} \includegraphics[width=\textwidth]{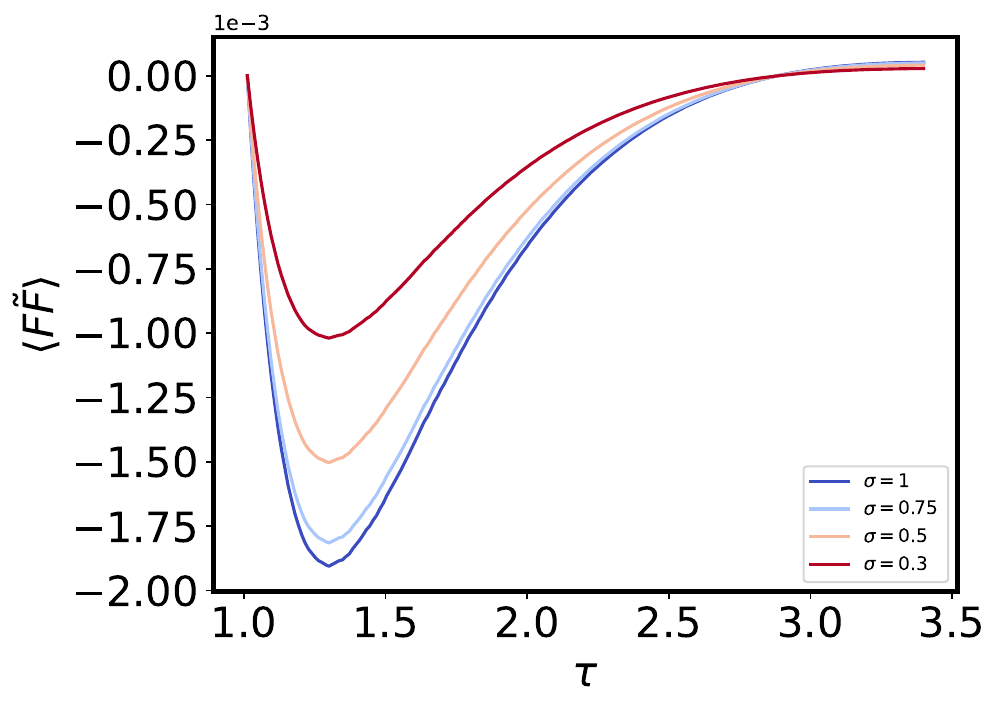}
\end{subfigure}
\caption{Evolution of the spatially averaged CS term. The left panel shows cases with different $\lambda_B$ and $\sigma_M=1$. The right panel shows cases with different $\sigma_M$ and $\lambda_B\sim0.2H_i^{-1}$.}
\label{fig:CS_term(10.30)}
\end{figure}

\begin{figure}[!htp]
\centering
\begin{subfigure}{0.4\textwidth} \includegraphics[width=\textwidth]{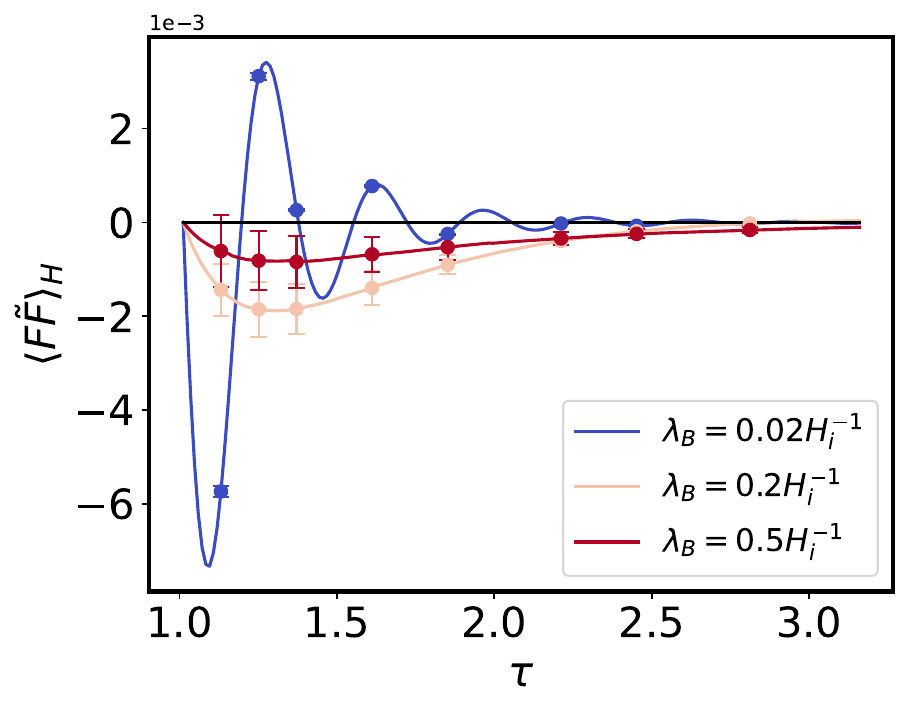}
\end{subfigure}
\begin{subfigure}{0.4\textwidth} \includegraphics[width=\textwidth]{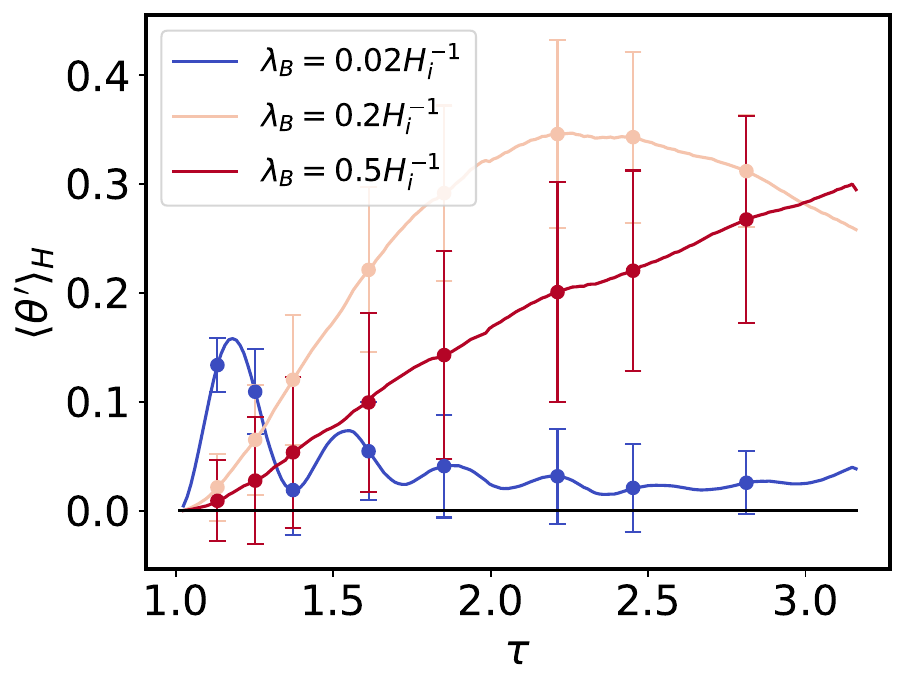}
\end{subfigure}
\caption{
Left: Time evolution of the Hubble-patch averaged CS term for three representative correlation lengths, $\lambda_B=0.02H_i^{-1}$, $0.2H_i^{-1}$ and $0.5H_i^{-1}$. At each time, the marker denotes the mean value obtained from all Hubble patches, while the error bar indicates the spread among different Hubble patches. Right: Same as the left panel, but for the Hubble-patch averaged scalar velocity $\langle\theta^{\prime}\rangle_H$. The horizontal black line denotes zero, serving as a reference for the sign of the Hubble-patch averaged quantities.
}
\label{fig:hubble_mean_coherence}
\end{figure}

For a CS source characterized by the wavenumber $k_{CS}$, the corresponding oscillation period is
\begin{equation}
    \Delta\tau_{\mathrm{CS}}
    \simeq
    \frac{\pi}{k_{\mathrm{CS}}}.
    \label{eq:CS_period}
\end{equation}
For a sufficiently narrow spectrum, $k_{\mathrm{CS}}\simeq\lambda_B^{-1}$, and hence
\begin{equation}
    \Delta\tau_{\mathrm{CS}}
    \simeq
    \pi\lambda_B.
    \label{eq:period_lambda}
\end{equation}
For the correlation lengths
\begin{equation}
    \lambda_BH_i
    =
    0.02,\ 0.05,\ 0.10,\ 0.20,
\end{equation}
the corresponding dimensionless oscillation timescales are
\begin{equation}
    \Delta\widetilde{\tau}_{\mathrm{CS}}
    \simeq
    0.38,\ 0.94,\ 1.88,\ 3.77
\end{equation}
respectively. These free-field estimates neglect axion backreaction and are intended only to capture the characteristic oscillation frequency inherited from the initial gauge-field spectrum. They capture the correlation-length dependence and the approximate oscillation periods observed in the left panels of Fig.~\ref{fig:CS_term(10.30)} and Fig.~\ref{fig:hubble_mean_coherence}, rather than providing an exact description of the full coupled evolution. Axion backreaction and nonlinear effects are fully retained in the lattice simulations.

\noindent{\it \bfseries The average CS source and the scaling parameter $\mathcal{A}$.} The CS interaction enters the axion equation of motion through the driving source $\mathcal{S}_{\rm CS}$ defined in Eq.~(7) of the main text. At the beginning of the simulation, its magnitude is large enough to affect the axion evolution, but it remains significant only during the short interval $1<\tau<2$. The response generated during this interval contains both an averaged component, which contributes to the bulk motion of the axion field, and an inhomogeneous component, which excites nonzero-momentum modes and modifies the local DW configuration.

We show the evolution of the average CS term at early times in the simulation in Fig.~\ref{fig:CS_term(10.30)}. The left panel presents results for different $\lambda_B$ with fixed $\sigma_M = 1$, while the right panel shows results for different $\sigma_M$ with fixed $\lambda_B \sim 0.2 H_i^{-1}$. For short correlation lengths, the averaged CS term oscillates rapidly, whereas its amplitude decreases as the correlation length increases. This behavior determines the accumulated bulk impulse of the axion field and contributes, together with the spatial fluctuations of the CS source, to the correlation-length dependence of the DW decay rate discussed in the main text. Furthermore, for a fixed correlation length, a decrease in $\sigma_M$ reduces the CS term, which in turn slows down the decay of the DW network. Since the CS term decreases as $a^{-4}$, we only plot its evolution during the very early stages.

\begin{figure}[!htp]
\begin{subfigure}{0.4\textwidth}
\includegraphics[width=\textwidth]{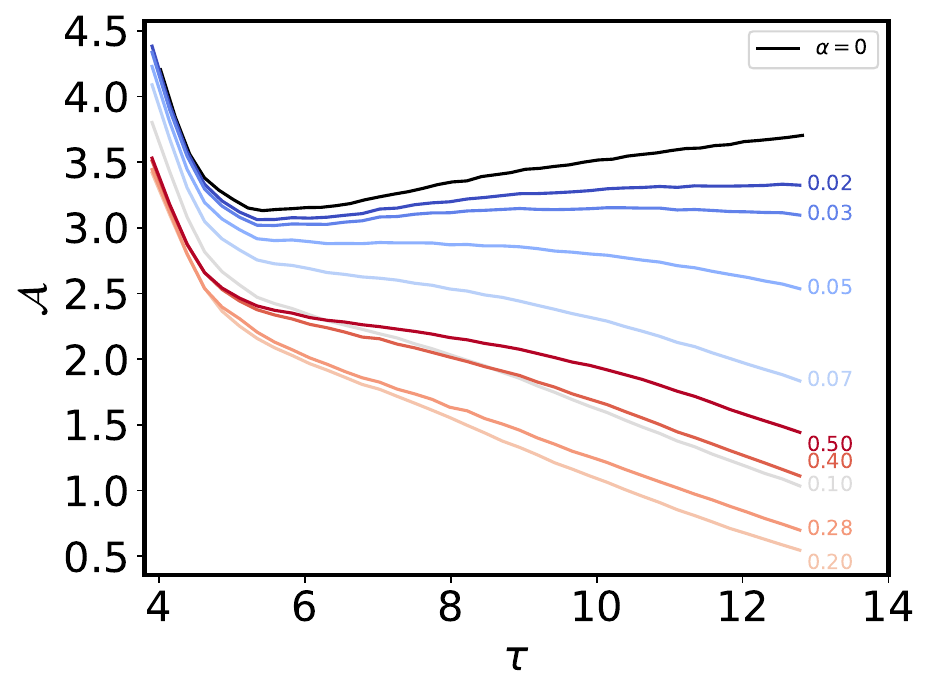}
\end{subfigure}
\begin{subfigure}{0.4\textwidth}
\includegraphics[width=\textwidth]{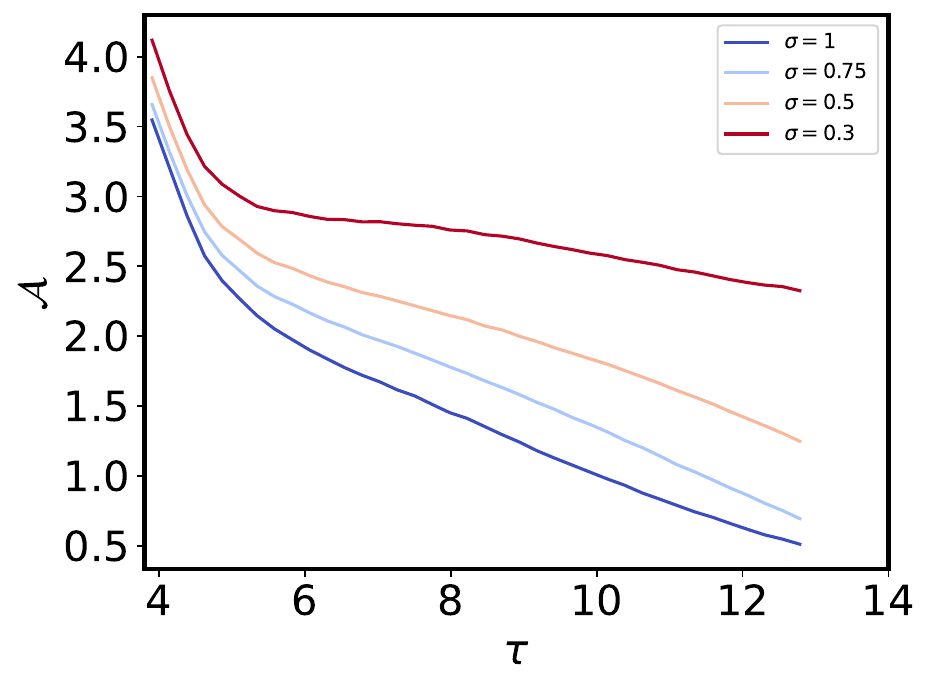}
\end{subfigure}
\caption{Evolution of the scaling parameter $\mathcal{A}$. The left panel shows cases with fixed $\sigma_M=1$. The value at the tail of each curve in the first figure indicates the corresponding $\lambda_B H_i$ associated with that case. The black curve corresponds to the case $\alpha=0$. The right panel shows cases with the fixed $\lambda_B\sim0.2H^{-1}_i$.}
\label{fig:DWs_scaling(10.30)}
\end{figure}

We then present the evolution of the scaling parameter in Fig.~\ref{fig:DWs_scaling(10.30)}. The left panel shows the results for different correlation lengths. For $\alpha = 0$, the DW network remains approximately in the scaling regime, whereas the decrease of $\mathcal{A}$ in the other cases signals the decay of DWs. The right panel of Fig.\ref{fig:DWs_scaling(10.30)} displays the scaling parameter for the case $\lambda_B \sim 0.2H_i^{-1}$ with different values of $\sigma_M$, which sets the helicity amplitude for magnetic fields with fixed energy density. As a result, $\sigma_M$ also controls the strength of the average CS term. Accordingly, the DW network decays more slowly for smaller $\sigma_M$. Notably, when $\sigma_M = 0.1$, the scaling parameter $\mathcal{A}$ evolves in nearly the same manner as in the $\alpha=0$ case, indicating that the helicity is too weak to affect the DW network evolution; therefore, we exclude this case from the subsequent discussion.

\begin{figure}[!htp]
\centering
\begin{subfigure}{0.4\textwidth} \includegraphics[width=\textwidth]{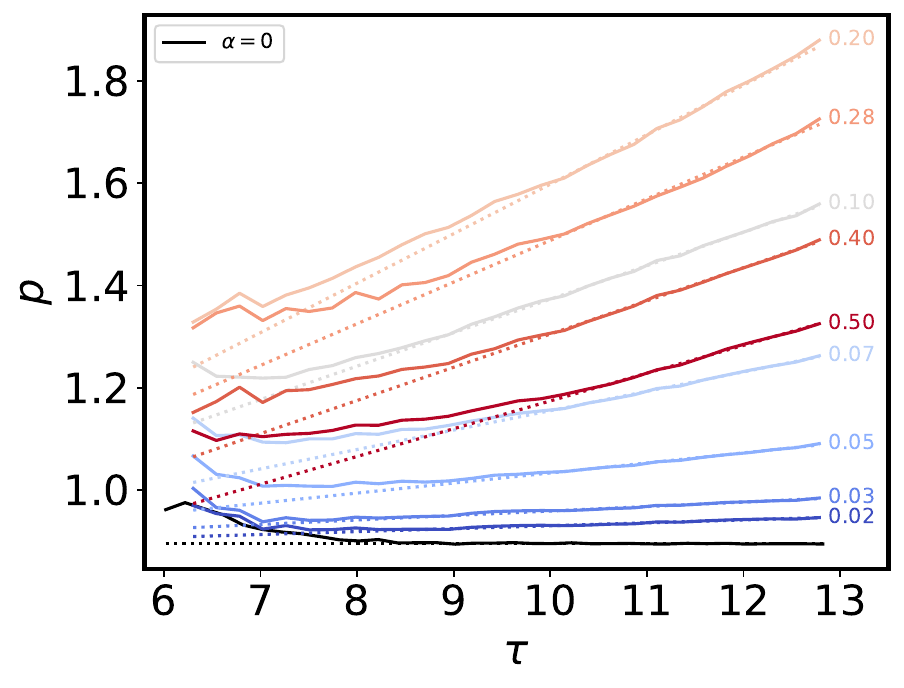}
\end{subfigure}
\begin{subfigure}{0.4\textwidth} \includegraphics[width=\textwidth]{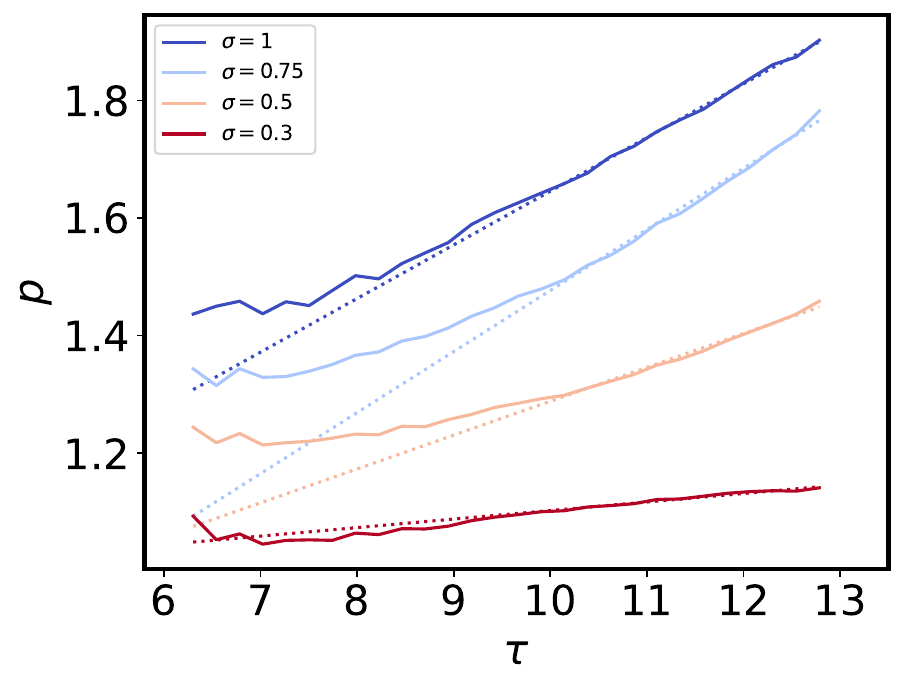}
\end{subfigure}
\caption{The evolution of $p$. The left panel shows cases with different $\lambda_B$ and $\sigma_M=1$, where the value at the tail of each curve indicates the corresponding $\lambda_B H_i$ associated with that case. The black curve corresponds to the case $\alpha = 0$. The right panel is cases with different $\sigma_M$ and $\lambda_B\sim0.2H_i^{-1}$. The dotted line shows the corresponding linear fit.}
\label{fig:p_evolve(10.30)}
\end{figure}
Finally, we use Eq.10 in the main text to show the evolution of the parameter $p$ in Fig.\ref{fig:p_evolve(10.30)}. The left panel presents results for different $\lambda_B$ with fixed $\sigma_M = 1$, while the right panel shows results for different $\sigma_M$ with fixed $\lambda_B \sim 0.2 H_i^{-1}$. In all cases, $p$ increases approximately linearly with time, and the dotted lines show the corresponding linear fits. Using this fitting function, we can determine the time at which $\mathcal{A}$ decreases to $10^{-2}$ or $10^{-1}$ of its value at the time of DW network formation, which defines the decay time. In addition, we define $t_{\rm form}$ as the time when the uncoupled ($\alpha=0$) network is close to entering the scaling regime and the value of $\mathcal{A}_{\rm form}$ is also taken at this time. The reason is that the initial correlated random field already crosses multiple vacua and generates spurious wall-like structures, so the true onset of physical DW formation cannot be cleanly identified. Since $t_{\rm form}$ and $A_{\rm form}$ are introduced at the same reference time, this choice does not affect the final results.

\noindent{\it \bfseries Hubble-scale coherence and local spatial structure of the CS-induced response.} We now examine the spatial structure of the CS source on both Hubble and sub-Hubble scales. At each time step, we divide the entire simulation volume into non-overlapping cubic subvolumes whose physical volume is of order one Hubble volume. Since the Hubble length evolves with time, the number of such subvolumes also changes during the evolution. In the early time interval considered in this coherence analysis, the number of Hubble-sized subvolumes in our simulation box decreases from about $700$ to about $60$. Within each Hubble volume, we compute the spatially averaged CS term $\langle F_{\mu\nu}\tilde F^{\mu\nu}\rangle_H $ and axion field velocity $ \langle\theta^{\prime}\rangle_H$.

The results are shown in Fig.~\ref{fig:hubble_mean_coherence}. The left panel presents the Hubble-patch averaged CS source for three representative magnetic-field correlation lengths. During the early evolution, the distributions contain a common large-scale component across the Hubble patches, although an appreciable patch-to-patch spread remains. This indicates that the nonzero volume average contains a coherent coarse-grained contribution, rather than being generated solely by finite-volume fluctuations of the patch-averaged source. At the same time, this large-scale component should not be interpreted as an energy splitting between different axion vacua. It instead characterizes the coarse-grained part of the dynamical source acting on the axion field.

The right panel of Fig.~\ref{fig:hubble_mean_coherence} shows the same analysis for the axion field velocity $\theta^{\prime}$. Compared with the CS term, the distribution of the Hubble-patch averaged velocity is visibly broader. This is expected. The scalar velocity is not determined solely by the instantaneous CS source, but instead reflects the accumulated response of the axion field to the full equation of motion, including the potential term and spatial-gradient contributions. Therefore, the Hubble-patch averaged $\theta^{\prime}$ need not be narrowly distributed around a single value. 

The large-scale coherence seen in Fig.~\ref{fig:hubble_mean_coherence} does not require the local CS source and the axion velocity to be aligned with each other point by point. Rather, it refers only to the presence of a coarse-grained component in the Hubble-patch averaged quantities. Local fluctuations around this component remain important because they excite nonzero-momentum axion modes and generate differential evolution among neighboring regions.

\begin{figure}[t]
\centering
\includegraphics[width=0.6\textwidth]{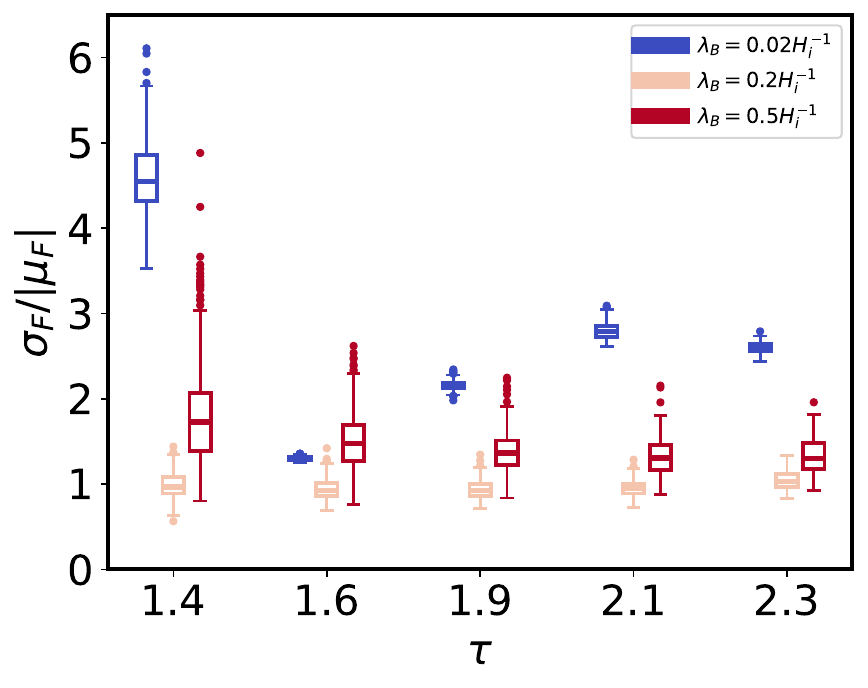}
\caption{
Box plots of the ratio $\sigma_F / |\mu_F|$ for the CS term inside each Hubble patch at several representative times, for $\lambda_B=0.02H_i^{-1}$, $0.2H_i^{-1}$ and $0.5H_i^{-1}$. Here $\mu_F$ and $\sigma_F$ denote respectively the mean value and standard deviation of $F_{\mu\nu}\tilde F^{\mu\nu}$ within one Hubble patch.
}
\label{fig:cs_intra_patch_variance}
\end{figure}

To quantify the spatial variation retained within each Hubble patch, we define $\mu_F$ and $\sigma_F$ as the mean and standard deviation of $F_{\mu\nu}\widetilde{F}^{\mu\nu}$ inside that patch. Figure~\ref{fig:cs_intra_patch_variance} presents box plots of the ratio $\sigma_F/|\mu_F|$. In each box plot, the central line denotes the median, the box spans the interquartile range, the whiskers indicate the spread of the main distribution, and the isolated points represent more extreme values. We find that $\sigma_F/|\mu_F|$ is generally of order unity or larger, with $\sigma_F$ reaching several times $|\mu_F|$ in some patches. The CS source is therefore strongly inhomogeneous even within a causally connected Hubble region.

Figures~\ref{fig:hubble_mean_coherence} and \ref{fig:cs_intra_patch_variance} thus demonstrate the coexistence of two spatial components of the CS source: a coherent coarse-grained component and sizable local fluctuations. The former contributes to the bulk displacement of the axion field, whereas the latter excites nonzero-momentum modes and produces differential evolution across the DW network. Neither the coarse-grained component nor the local fluctuations introduce a preferred vacuum or lift the degeneracy of the $Z_{N_{\rm DW}}$ potential.

\noindent\textit{\textbf{Local thin-wall interpretation.}} In a sufficiently small neighborhood of a smooth wall, choose a reference plane with coordinates $\mathbf{x}_{\parallel}$ and unit normal $\hat{\mathbf n}$. Points in this neighborhood are written as $\mathbf{x}=\mathbf{x}_{\parallel}+\xi\hat{\mathbf n}$, while the wall is represented by the height function $\xi=q(\mathbf{x}_{\parallel},\tau)$. The variable
\begin{equation}
    u\equiv \xi-q(\mathbf{x}_{\parallel},\tau)
\end{equation}
is therefore the normal displacement from the wall within this local description. We decompose the axion field as
\begin{equation}
    \phi(\mathbf{x},\tau) =\bar\phi(\tau)+\phi_w(u)+\delta\phi(\mathbf{x},\tau), 
    \label{eq:local-wall-decomposition}
\end{equation}
where $\bar\phi$ is the spatially averaged background field, $\phi_w$ is the wall profile measured relative to this background, and $\delta\phi$ contains the remaining fluctuations. For a wall connecting adjacent vacua,
\begin{equation}
    \Delta\phi \equiv \phi_w(+\infty)-\phi_w(-\infty) =\pm\frac{2\pi v}{N_{\rm DW}},
    \label{eq:wall-excursion}
\end{equation}
where the sign specifies the orientation of the wall.

To see how the wall position enters the field equation, we first neglect $\delta\phi$ and differentiate Eq.~\eqref{eq:local-wall-decomposition}. Here and below, primes on $q$ and $\bar\phi$ denote conformal-time derivatives, while derivatives of the wall profile are written explicitly as $\partial_u$. We obtain
\begin{align}
    \partial_\tau\phi &=\bar\phi'-q'\partial_u\phi_w, \\ 
    \partial_\tau^2\phi &=\bar\phi''-q''\partial_u\phi_w +(q')^2\partial_u^2\phi_w, \\ \nabla^2\phi 
    &=\left[1+(\boldsymbol{\nabla}_{\parallel}q)^2\right] \partial_u^2\phi_w -(\nabla_{\parallel}^2q)\partial_u\phi_w .
    \label{eq:wall-profile-derivatives}
\end{align}
Here $\boldsymbol{\nabla}_{\parallel}$ is the gradient operator with respect to the coordinates $\mathbf{x}_{\parallel}$ on the reference plane. The axion equation, with the CS source decomposed into its spatially
averaged and inhomogeneous components, is
\begin{equation}
    \phi''-\nabla^2\phi+2\mathcal H\phi' +a^2V_{\phi}=S_{\rm CS}, \qquad S_{\rm CS}=\overline{S}_{\rm CS}+\delta S_{\rm CS},
    \label{eq:axion-local-source}
\end{equation}
Subtracting the spatially averaged equation from Eq.~\ref{eq:axion-local-source} gives
\begin{align}
    &\left[(q')^2-(\boldsymbol{\nabla}_{\parallel}q)^2-1\right] \partial_u^2\phi_w +\left(-q''+\nabla_{\parallel}^2q-2\mathcal Hq'\right) \partial_u\phi_w +\mathcal R =\delta S_{\rm CS},
    \label{eq:local-wall-field-equation}
\end{align}
where $\mathcal R$ collects the potential terms, deformations of the ideal wall profile, and terms involving $\delta\phi$. 

The direct force can be identified without introducing a local vacuum-energy difference. Under an infinitesimal wall displacement $\delta q$, the profile changes by
\begin{equation}
    \delta_q\phi_w=-\delta q\,\partial_u\phi_w .
\end{equation}
Since $\delta S_{\rm CS}$ appears as a source in the field equation, it acts as a generalized force density conjugate to the field $\phi$. The corresponding virtual work per unit area is obtained by integrating $\delta S_{\rm CS}\delta_q\phi_w$ across the wall core:
\begin{equation}
    \delta W_{\rm CS} =\int du\,\delta S_{\rm CS}\,\delta_q\phi_w =F_{\rm CS}\,\delta q ,
\end{equation}
with
\begin{equation}
    F_{\rm CS}(\mathbf{x}_w,\tau) =-\int du\,\partial_u\phi_w(u)\, \delta S_{\rm CS} (\mathbf{x}_w+u\hat{\mathbf n},\tau), \qquad \mathbf{x}_w =\mathbf{x}_{\parallel}+q\hat{\mathbf n}.
    \label{eq:cs-relative-force}
\end{equation}
Thus, $F_{\rm CS}$ is the generalized force per unit area conjugate to the collective coordinate $q$, and it directly drives the normal motion of the local wall segment. In particular,
\begin{equation}
    \int du\,\partial_u\phi_w\,\partial_u^2\phi_w =\frac{1}{2} \left[(\partial_u\phi_w)^2\right]_{-\infty}^{+\infty} =0,
\end{equation}
because $\partial_u\phi_w$ vanishes outside the wall core. Defining the profile normalization
\begin{equation}
    \Sigma_w\equiv\int du\,(\partial_u\phi_w)^2,
\end{equation} the leading collective-coordinate equation is therefore
\begin{equation}
    \Sigma_w \left(q''+2\mathcal Hq'-\nabla_{\parallel}^2q\right) =F_{\rm CS}+\cdots ,
    \label{eq:collective-wall-equation}
\end{equation}
where the three terms on the left describe wall inertia, Hubble damping, and curvature, respectively. The ellipsis denotes the projected terms in $\mathcal R$ and corrections to the local height-function approximation.

If the CS source varies slowly across the wall core, such that its characteristic length scale $\lambda_{CS}$ is much larger than the DW width, $r_{\rm DW}\ll \lambda_{\rm CS}$, it can be expanded about the wall center.
\begin{equation}
    \delta S_{\rm CS} (\mathbf{x}_w+u\hat{\mathbf n},\tau) = \delta S_{\rm CS}(\mathbf{x}_w,\tau) +u\,\partial_n\delta S_{\rm CS}(\mathbf{x}_w,\tau)
    +\cdots .
\end{equation}
Using $\int du\,\partial_u\phi_w=\Delta\phi$, the leading term becomes
\begin{equation}
    F_{\rm CS}(\mathbf{x}_w,\tau) \simeq -\Delta\phi\,\delta S_{\rm CS}(\mathbf{x}_w,\tau).
    \label{eq:thin-wall-local-force}
\end{equation}
Thus, spatial variations of the CS source give different normal accelerations to different wall segments. Reversing the wall orientation changes the sign of $\Delta\phi$, while $\delta S_{\rm CS}$ itself varies across the network. The resulting force therefore does not select a globally preferred vacuum or lift the degeneracy of the $Z_{N_{\rm DW}}$ potential. Instead, it changes the wall shape, curvature, and collision history, thereby facilitating the reduction of the total wall area. The averaged source can still influence the network indirectly through $\bar\phi(\tau)$ and the nonlinear potential terms.

The condition $r_{\rm DW}\ll\lambda_{CS}$ is required only for the local approximation in Eq.~\ref{eq:thin-wall-local-force}. When the two scales are not well separated, the source must be retained inside the profile integral, and wall deformation and internal gradients can become important. These effects are automatically included in the full lattice evolution, which does not rely on the thin-wall approximation. In addition, an appreciable differential response across the network requires the CS source to vary on scales not much larger than the characteristic wall separation or curvature scale.

\begin{figure}[!htp]
\centering
\begin{subfigure}{0.4\textwidth} \includegraphics[width=\textwidth]{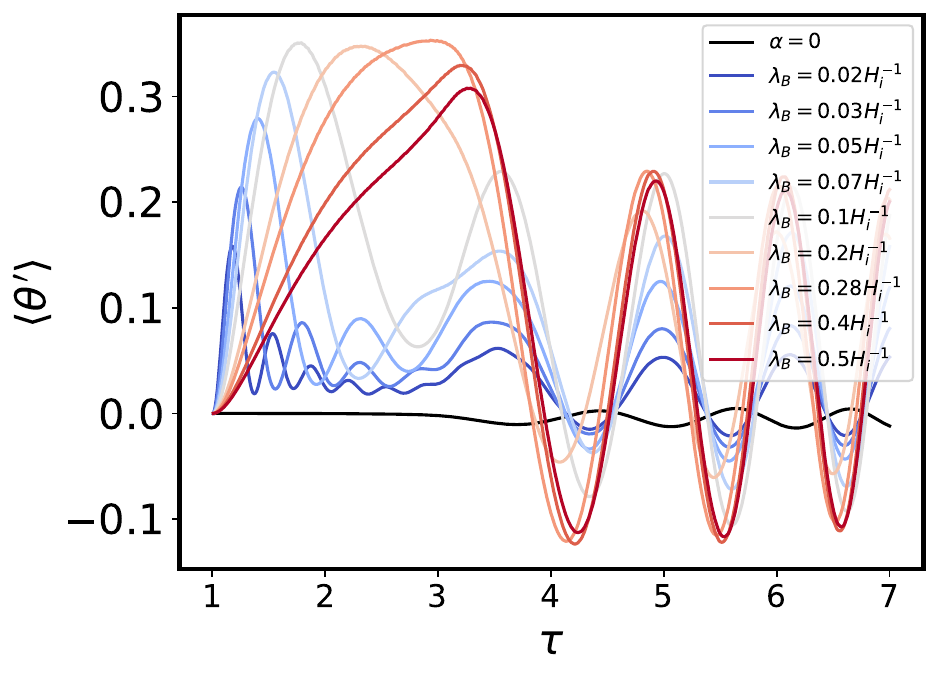}
\end{subfigure}
\begin{subfigure}{0.39\textwidth} \includegraphics[width=\textwidth]{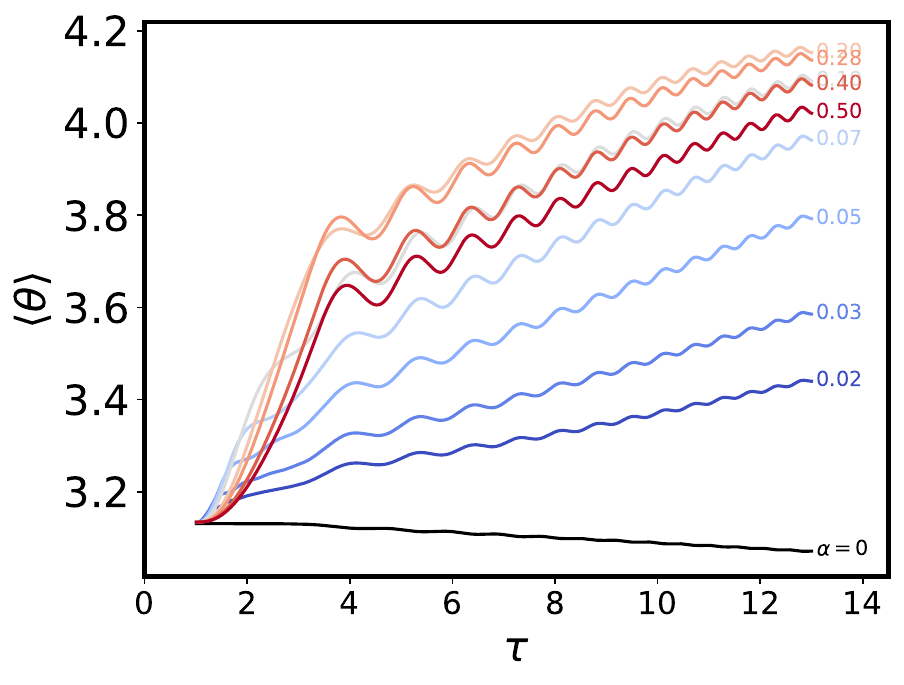}
\end{subfigure}
\caption{
Left: Time evolution of the average velocity of axion field $\theta^{\prime}$ in the early time. Right: Time evolution of the averaged field $\theta$ for different magnetic-field correlation lengths.
}
\label{fig:avg_theta}
\end{figure}

\noindent{\it \bfseries Axion response and the collapse of DWs.} As discussed in the main text, the CS source affects both the spatially averaged axion mode and the inhomogeneous field configuration. Fig.~\ref{fig:avg_theta} characterizes the averaged response generated during the early stage of the evolution.

The left panel of Fig.~\ref{fig:avg_theta} shows the spatially averaged scalar velocity $\langle\theta'\rangle$. In the uncoupled case $\alpha=0$, this quantity remains close to zero throughout the evolution. Once the CS coupling is turned on, a nonzero averaged velocity is generated at early times and remains visible around $\tau\sim4$, when the uncoupled DW network is approaching its scaling regime. This demonstrates that the averaged CS source produces a genuine bulk
response of the axion field.

The right panel shows the corresponding evolution of the averaged field $\langle\theta\rangle$. The early nonzero velocity accumulates into a finite displacement that remains visible after the CS source has become small. The two panels therefore characterize the homogeneous part of the axion response. The evolution of the DW network additionally depends on the nonzero-momentum modes excited by the spatially varying CS source. These modes drive differential motion of neighboring wall segments and continuously rearrange the local domain configuration.

To further clarify why the CS term can dynamically affect DW formation, we compare the velocity induced by the CS source with the two characteristic velocity scales associated with the Hubble time and the axion oscillation time. From the axion equation of motion, the CS contribution to the dimensionless acceleration of $\theta$ is
\begin{equation}
    S_{\rm CS}=\left|\frac{\alpha a^2}{4w_*^2 v^2}\left\langle F_{\mu\nu}\tilde F^{\mu\nu}\right\rangle\right| .
\end{equation}
For the representative case $\lambda_B=0.2H_i^{-1}$, we find
$\left\langle F_{\mu\nu}\tilde F^{\mu\nu}\right\rangle\simeq 2\times 10^{-3}\,{\rm GeV}^{4}$ around the time of DW formation, identified by the condition $m_a=3H$. Multiplying this instantaneous CS acceleration by the two dimensionless characteristic conformal-time intervals, $\Delta \tilde{\tau}_H\simeq w_*/(aH)$ and $\Delta \tilde{\tau}_m\simeq w_*/(a m_a)$, and comparing the resulting velocity increments with the corresponding dimensionless characteristic velocities, $v_H\simeq (aH/w_*)\theta$ and $v_m\simeq (a m_a/w_*)\theta$, we obtain
\begin{equation}
    \frac{v_{\rm CS,H}}{v_H}
    =
    \frac{S_{\rm CS}\Delta \tilde{\tau}_H}{(aH/w_*)\theta}
    \simeq 21.4,
    \qquad
    \frac{v_{\rm CS,m}}{v_m}
    =
    \frac{S_{\rm CS}\Delta \tilde{\tau}_m}{(a m_a/w_*)\theta}
    \simeq 2.4 .
\end{equation}

This order-of-magnitude estimate shows that, in this representative case, the instantaneous CS source is dynamically important before and around the onset of DW formation. The estimate is based on the spatially averaged source and therefore characterizes the scale of the bulk axion impulse. In the full lattice simulation, however, the source oscillates in time and varies across space. The accumulated impulse is consequently distributed between the homogeneous mode and nonzero-momentum modes. The latter generate spatially varying axion responses that drive differential motion of local wall segments, change the relative volumes of neighboring domains, and contribute directly to the reduction of the DW area.

\begin{figure}[!htp]
\centering
\begin{subfigure}{0.6\textwidth} \includegraphics[width=\textwidth]{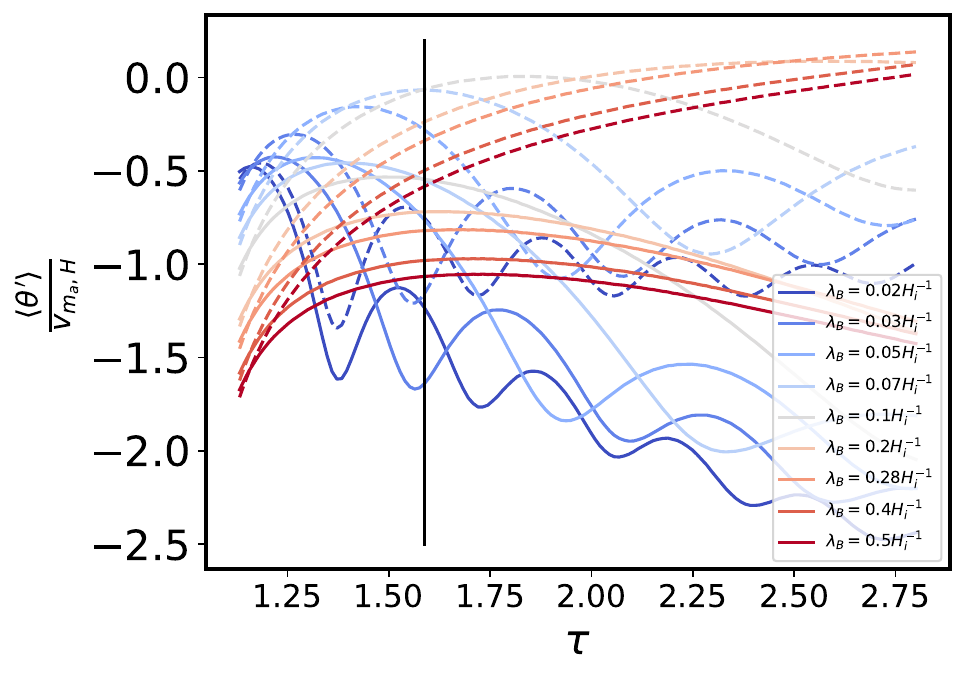}
\end{subfigure}
\caption{
The ratios $\langle \theta' \rangle / v_{m_a}$ and $\langle \theta' \rangle / v_H$ for each case at early times, where $v_{m_a}=a\frac{m_a}{w_*}\theta$ and $v_H=\frac{\mathcal{H}}{w_*}\theta = a\frac{H}{w_*}\theta$ are the two dimensionless characteristic velocities. The solid and dashed curves denote $\langle\theta'\rangle/v_{m_a}$ and $\langle\theta'\rangle/v_H$, respectively. The black line indicates the epoch $m_a = 3H$.
}
\label{fig:avg_theta2}
\end{figure}

Figure~\ref{fig:avg_theta2} provides a complementary diagnostic using the actual spatially averaged velocity measured in the full lattice simulation. We plot the signed ratios of $\langle\theta'\rangle$ to the two characteristic velocities defined above. These ratios vary appreciably among the different correlation lengths, and their magnitudes can reach order unity in some cases. Therefore, the retained homogeneous velocity is not uniformly negligible compared with both characteristic velocity scales. Nevertheless, it is generally reduced relative to the constant-source impulse estimate, with the reduction being particularly significant for the Hubble-time estimate.

This difference is expected because $S_{\rm CS}\Delta\widetilde{\tau}$ estimates the velocity that would be generated if the spatially averaged CS acceleration remained constant and coherent over one characteristic time scale. In the full simulation, $\langle F_{\mu\nu}\widetilde F^{\mu\nu}\rangle$ oscillates and redshifts rapidly, while the induced homogeneous momentum is damped by Hubble friction. Part of the CS impulse is also transferred to nonzero-momentum axion modes, and the resulting inhomogeneous field configuration feeds back on the averaged motion through the nonlinear term $\langle V'(\phi)\rangle$. Figure~\ref{fig:avg_theta2} therefore shows that the late-time evolution cannot be characterized by a persistent homogeneous velocity alone. The early CS impulse is retained partly as an averaged field displacement and partly as an inhomogeneous axion response that deforms the DW network.

\noindent{\it \bfseries The calculation of GW.} The GW is generated through oscillations of DWs. In lattice simulations, the evolution of the metric perturbation $h_{ij}$ is governed by
\begin{equation}
\label{eq:eom_of_hij}
h^{\prime\prime}_{ij}-\nabla^2 h_{i j}+2\mathcal{H}{h}^{\prime}_{ij}=16 \pi G T_{ij}^{\mathrm{TT}},
\end{equation}
where $T^{\mathrm{TT}}_{ij}$ is the transverse-traceless component of the energy-momentum tensor, given by \cite{pystella}:
\begin{align}
\label{eq:energy-momentum_tensor}
T_{ij}^{\mathrm{TT}}&=\partial_i\phi\partial_j\phi+F_{i\alpha}F_{j\beta }g^{\alpha\beta},
\end{align}
with $g_{\mu\nu}$ being the FLRW metric. The dimensionless GW spectrum is then calculated by:
\begin{equation}\label{eq:GW_spectrum_lattice}
\Omega_{gw}(k)=\frac{1}{\rho_c}\frac{\partial \rho_{gw}(k)}{\partial \ln{k}}=\frac{1}{24\pi^2 V} \frac{k^3}{\mathcal{H}^2}\sum_{ij}|h_{ij}^{\prime}(k,\tau)|^2,
\end{equation}
where $h_{ij}(k,\tau)$ denotes the Fourier transform of $h_{ij}(x)$, $V$ is the comoving volume of the whole lattice and $\rho_c=\frac{3}{8\pi G}H^2$ is the critical energy density. When calculating the GW spectrum from the lattice simulation, we use the Runge-Kutta method to numerically solve Eq.\eqref{eq:eom_of_hij}, and the calculation of the GW spectrum within the lattice simulation is performed using the Pystella code introduced in Ref.~\cite{pystella}.

As discussed in the text, the peak amplitude of GW spectra can be parameterized as $\Omega_{gw}^{p}\propto H^{-2}f_a^4m_a^2$.
\begin{align}
\label{eq:Omega_peak}
\Omega_{gw}^{p}(t) &=
\tilde{\Omega}_1 \frac{f_a^4m_a^2}{H(t)^2}\sim
\tilde{\Omega}\left(\frac{t_{\text{dec}}}{t_{\text{form}}}\right)^2 \left(\frac{f_a}{1.8\times10^{16}\text{GeV}}\right)^4,
\end{align}
where we have used $H\sim m_a$ when the DW network forms and $H=1/(2t)$ in the radiation-dominated Universe. Here, the parameters $\tilde{\Omega}_1$ and $\tilde{\Omega}$ are two distinct proportionality coefficients. We stress that the $t_{\rm form}$ used in Eq.10 in the main text is an operational time, defined near the onset of the scaling regime, rather than the true physical formation time of the wall network. Hence $m_a \gg H$ at $t_{\rm form}$, and the approximation $H\sim m_a$ mentioned before is only an overall normalization difference, which can be absorbed into the fitted coefficient $\tilde{\Omega}$. In addition, $t_{\text{dec}}$ denotes the operational decay time defined by the wall-area criterion specified in the main text. In the second expression, we normalize to our simulation value $f_a=1.8\times10^{16}\mathrm{GeV}$.

Using entropy conservation, $T\propto g_{S}(T)^{-1/3}a^{-1}$, and $\rho_{\rm gw}\propto a^{-4}$, we obtain the present-day peak amplitude
\begin{equation}
\Omega_{gw,0}^{p}=\tilde{\Omega} \left(\frac{t_{\text{dec}}}{t_{\text{form}}}\right)^2 \left(\frac{f_a}{1.8\times10^{16}\text{GeV}}\right)^4 \left( \frac{10.75}{g_*(T_{\text{dec}})} \right)^{{4/3}},
\end{equation}
where we have used $g_{S}(T)\approx g_*(T)$.

\begin{figure}[!htp]
\centering
\begin{subfigure}{0.4\textwidth}
\includegraphics[width=\textwidth]{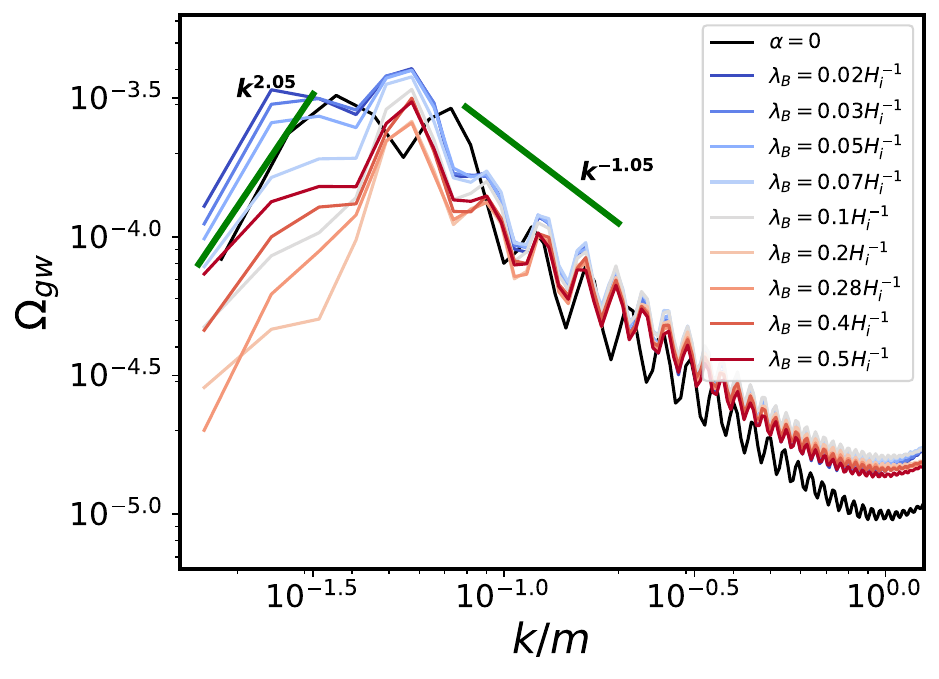}
\end{subfigure}
\begin{subfigure}{0.375\textwidth}
\includegraphics[width=\textwidth]{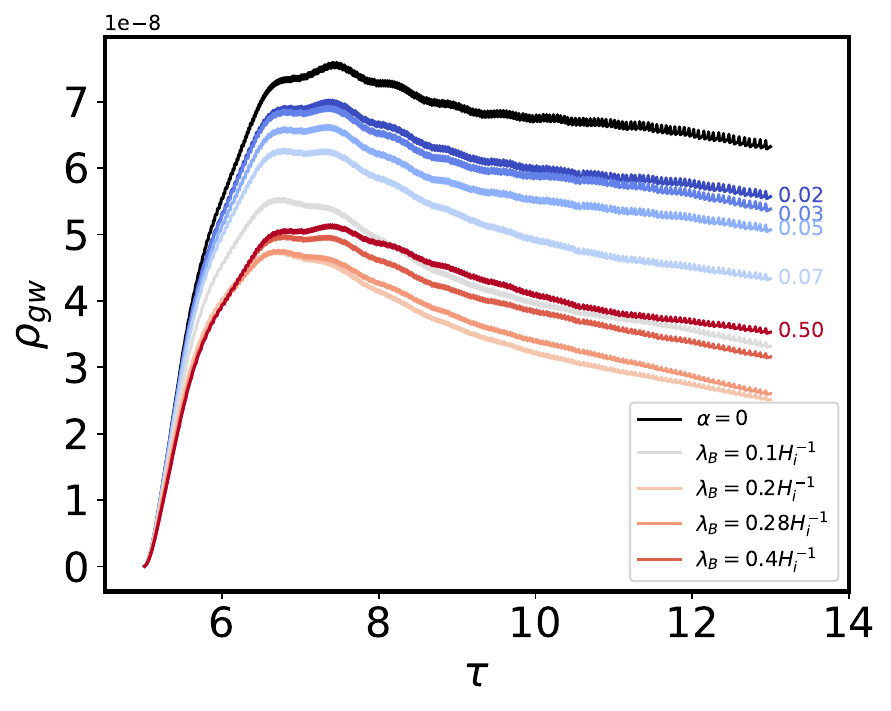}
\end{subfigure}
\caption{Left: The GW spectra at $\tau = 10$ are shown. The unphysical high-frequency region corresponding to scales smaller than the DW width (with $k > m_a$) has been removed. Right: Evolution of the GW energy density. The values at the tail of each curve indicate the corresponding $\lambda_BH_i$ associated with that case. The black curve corresponds to the case $\alpha=0$. Several curves are too close to label at the curve ends; we therefore indicate the corresponding correlation lengths in the legend.}
\label{fig:GW_spectra(10.30)}
\end{figure}

We present the GW spectra emitted at $\tau = 10$ for each case in Fig.~\ref{fig:GW_spectra(10.30)}. The low-frequency part of the spectra is small for fast-decay cases, because the decay of DWs reduces the GW source, particularly at large scales. Near the left peak of the GW spectra, these curves follow power-law dependencies $k^{2.05}$ in the low-frequency regime and $k^{-1.05}$ in the
mid-frequency regime. Then we show the evolution of the GW energy density for each case in Fig.~\ref{fig:GW_spectra(10.30)}. In the case without coupling, the GW energy density approaches a constant value, indicating the onset of the scaling regime \cite{2406.17053}. In contrast, when a fully helical magnetic field is included, the GW energy density decreases as the DW network decays.

\noindent{\it \bfseries The DW-radiated axion energy density.} We focus exclusively on the gauge-coupling cases. Assuming that all energy emitted by the DW network is transferred to free axions \cite{1412.0789}, 
\begin{align}
    \frac{d\rho_{dw}}{dt}=-H\rho_{dw}-\frac{d\rho_{dw}}{dt}\Big|_{e} \notag \\
    \frac{d\rho_{a}}{dt}=-3H\rho_{a}+\frac{d\rho_{dw}}{dt}\Big|_{e}
\end{align}
Strictly speaking, different momentum modes for axion field redshift differently, so a fully accurate treatment would require integrating over the whole momentum distribution. Here we adopt the same effective approximation as Ref.~\cite{1412.0789} and treat the emitted axions as approximately nonrelativistic. Using the DW energy density $\rho_{dw}=\frac{\mathcal{A}(t)\sigma_{dw}}{t}$, where $\mathcal{A}(t)$ is determined by Eq.10 in the main text, and the time-dependent parameter $p=p(t)$, we can get 
\begin{equation}
    \frac{d\rho_{dw}}{dt}\Big|_{e}=\frac{2p(t)-1}{2t} \rho_{dw}-\frac{dp(t)}{dt} \ln{\Big(\frac{t_f}{t}}\Big) \rho_{dw}
\end{equation}
Then the energy density of axion is 
\begin{align}
    a^3(t_{\text{dec}})\rho_{a}(t_{\text{dec}})&=\int_{t_{\text{form}}}^{t_{\text{dec}}} dt^{\prime} a^3(t^{\prime})\left[ \frac{2p(t^{\prime})-1}{2t^{\prime 2}} \mathcal{A}(t^{\prime})\sigma_{dw} \right] - \int_{t_{\text{form}}}^{t_{\text{dec}}} dt^{\prime} a^3(t^{\prime})\left[ \frac{dp(t^{\prime})}{dt^{\prime}} \ln \left( \frac{t_f}{t^{\prime}}\right) \frac{\mathcal{A}(t^{\prime})\sigma_{dw}}{t^{\prime}} \right] \notag \\
    &=\mathcal{A}_{\text{form}}\sigma_{dw}\int_{t_{\text{form}}}^{t_{\text{dec}}} dt^{\prime} a^3(t^{\prime})\left[ \frac{2p(t^{\prime})-1}{2t^{\prime 2}}-\frac{dp(t^{\prime})}{dt^{\prime}} \ln \left( \frac{t_f}{t^{\prime}}\right)  \frac{1}{t^{\prime}}\right] \left(\frac{t^{\prime}}{t_f}\right)^{1-p(t^{\prime})} \notag \\
    &=\mathcal{A}_{\text{form}}\sigma_{dw}\int_{t_{\text{form}}}^{t_{\text{dec}}} dt^{\prime} a^3(t^{\prime})\left[ \frac{2p(t^{\prime})-1}{2t^{\prime 2}}-\frac{k^{\prime}}{a(t^{\prime})t^{\prime}} \ln \left( \frac{t_f}{t^{\prime}}\right)\right] \left(\frac{t^{\prime}}{t_f}\right)^{1-p(t^{\prime})} \notag \\
    &=\frac{\mathcal{A}_{\text{form}}\sigma_{dw}}{t_\text{form}}\int_1^{s_{\text{dec}}} ds a^3(s)\left[ \frac{2p(s)-1}{2s^2}+\frac{k}{a(s)s} \ln {s}\right] s^{1-p(s)}
\end{align}
where we have used the fitting function $p(\tau)=k^{\prime}\tau+b$ and define the parameter $s=t/t_{\text{form}}, k=k^{\prime}t_{\text{form}}$.

Introducing the dimensionless efficiency factor $\kappa$, we obtain
\begin{equation}
    \rho_a(t_{\text{dec}})=\kappa \frac{\mathcal{A}_{\text{form}}\sigma_{dw}}{t_\text{form}},\quad \kappa=\int_1^{s_{\text{dec}}} ds \frac{a^3(s)}{a^3(s_{dec})}\left[ \frac{2p(s)-1}{2s^2}+\frac{k}{a(s)s} \ln {s}\right] s^{1-p(s)}
\end{equation}
Here $\frac{\mathcal{A}_{\text{form}}\sigma_{dw}}{t_\text{form}}$ represents the DW energy density at the formation time, while the dimensionless factor $\kappa$ parameterizes how the continuous emission history and the time dependence of $p(t)$ are accumulated into the final axion energy density under the assumption that the emitted DW energy is entirely converted into free axions.

Then we consider the redshift. Using the above result and the conservation of entropy,
\begin{align}
    \rho_{a}(t_{\text{dec}})=\kappa\frac{\mathcal{A_{\text{form}}}\sigma_{dw}}{t_{\text{form}}}=2\kappa\mathcal{A_{\text{form}}}\sigma_{dw}\left(\frac{t_{\text{dec}}}{t_{\text{form}}}\right)H_d, \qquad
    s=\frac{2\pi^2}{45}g_{S}(T)T^3\propto a^{-3}
\end{align}
we obtain the axion energy density after the DW network collapses
\begin{equation}
\rho_a(t)=m_a(0)n_a(t_{\text{dec}})\left(\frac{a(t_{\text{dec}})}{a(t)}\right)^3=2\frac{\kappa}{\epsilon_a}\mathcal{A_{\text{form}}}\sigma_{dw}\left(\frac{t_{\text{dec}}}{t_{\text{form}}}\right)H_d\left(\frac{s(t)}{s(t_{\text{dec}})}\right).
\end{equation}
Using $H_d^2=\frac{\pi^2}{90M_{pl}^2}g_*(T_\text{dec})T_{\text{dec}}^4$ and $g_*(T_{\text{dec}})\approx g_S(T_{\text{dec}})$, we obtain the present-day axion density parameter $\Omega_{a,0}=\rho_{a,0}/\rho_{c,0}:$
\begin{equation}
    \Omega_{a}\sim \frac{1}{\rho_{c,0}}\frac{\kappa}{\epsilon_a}\mathcal{A}_{\text{form}}\beta f_a^2m_a\left(\frac{t_{\text{dec}}}{t_{\text{form}}}\right)H_d\left(\frac{s_0}{s_{\text{dec}}}\right)
    \sim \frac{1}{\rho_{c,0}}\frac{g_S(T_0)T_0^3}{M_{pl}^{3/2}}\frac{\kappa}{\epsilon_a}\mathcal{A}_{\text{form}}\beta f_a^2m_a^{1/2}\left(\frac{t_{\text{dec}}}{t_{\text{form}}}\right)^{3/2} \left(\frac{1}{g_*(T_{\text{dec}})}\right)^{1/4} 
\end{equation}
where we have used $H\sim m_a$ when the DW network forms and $H=1/(2t)$. The parameter $\beta$ is defined as $\beta=\sigma_{dw}/(m_af_a^2)=9.32$ and $\epsilon_a=\omega_a/m_a$.

Substituting numerical constants, we obtain
\begin{align}
\label{eq:ratio_axion_DM}
\Omega_{a,0}
\sim &1.3\times10^5 \frac{\kappa}{\epsilon_a}\mathcal{A}_{\text{form}} \left(\frac{t_{\text{dec}}}{t_{\text{form}}}\right)^{3/2}\left(\frac{f_a}{1.8\times 10^{16}\text{GeV}}\right)^2 \left(\frac{m_a}{3.2\times 10^{-10}\text{eV}}\right)^{1/2} \left(\frac{10.75}{g_*(T_{\text{dec}})}\right)^{1/4}
\notag \\
\sim &0.0225 \frac{\kappa}{\epsilon_a}\mathcal{A}_{\text{form}} \left(\frac{t_{\text{dec}}}{t_{\text{form}}}\right)^{3/2}\left(\frac{f_a}{10^{12}\text{GeV}}\right)^2 \left(\frac{m_a}{ 10^{-6}\text{eV}}\right)^{1/2}\left(\frac{10.75}{g_*(T_{\text{dec}})}\right)^{1/4}.
\end{align}

The first line is normalized to the simulation parameters: $f_a=1.8\times10^{16}\text{GeV}, m_a=3.2\times10^{-10}\text{eV}$. The large coefficient $1.3\times10^5$ shows that these parameters are so large that the axion is over-produced. The last row is normalized by the parameters shown in Ref.~\cite{2212.13263}: $f_a\sim10^{12}\text{GeV}, m_a\sim10^{-6}\text{eV}$.

\begin{figure}[!htp]
\centering
\begin{subfigure}{0.39\textwidth}
\includegraphics[width=\textwidth]{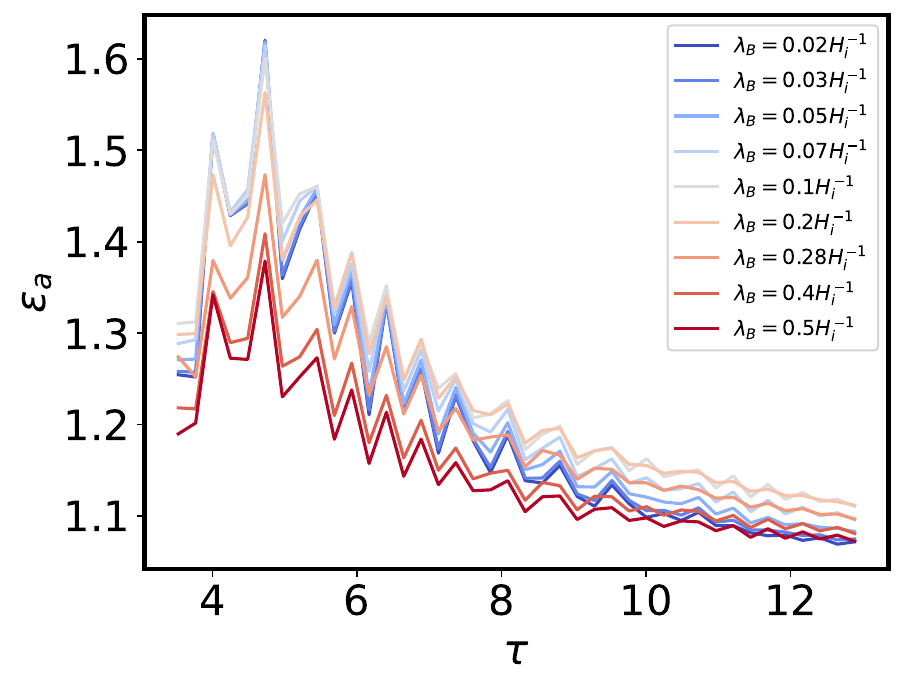}
\end{subfigure}
\begin{subfigure}{0.41\textwidth}
\includegraphics[width=\textwidth]{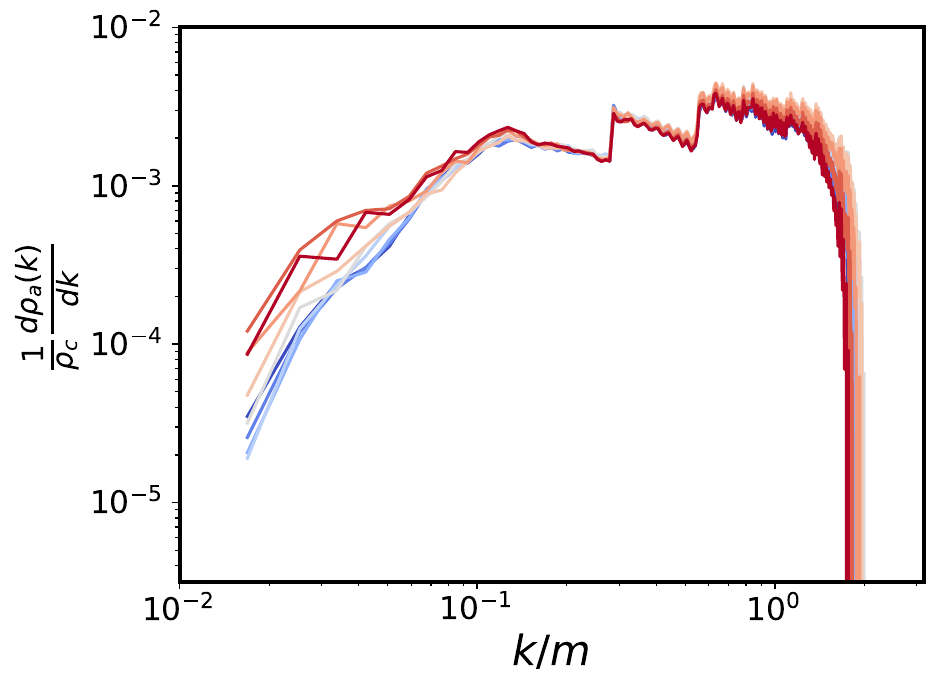}
\end{subfigure}
\caption{The average momentum parameter $\epsilon_a$ and axion energy density spectra at $\tau=11$. In the right panel, the correspondence between color and $\lambda_B$ is the same as the legend in the left panel.}
\label{fig:axion_spectra(10.30)}
\end{figure}

The above calculation requires the average momentum parameter $\epsilon_a$, which can be obtained from the spectrum of the axion energy density. We plot the evolution of $\epsilon_a$ in the left panel of Fig.~\ref{fig:axion_spectra(10.30)}, and the right panel shows the spectrum radiated at $\tau = 11$. In the spectra figure, we remove the unphysical high-frequency component by subtracting the initial spectrum. The spectra are all peaked in the high-frequency region, and are larger for the long-correlation cases, which is consistent with the results of $\epsilon_a$.

\begin{figure}[!htp]
\centering
\begin{subfigure}{0.4\textwidth}
\includegraphics[width=\textwidth]{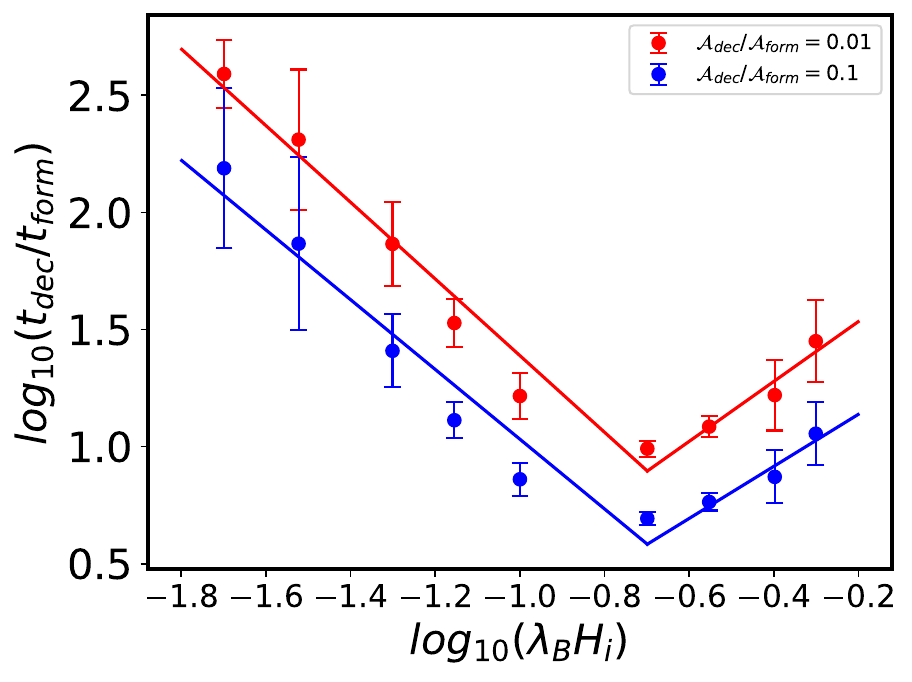}
\end{subfigure}
\begin{subfigure}{0.4\textwidth}
\includegraphics[width=\textwidth]{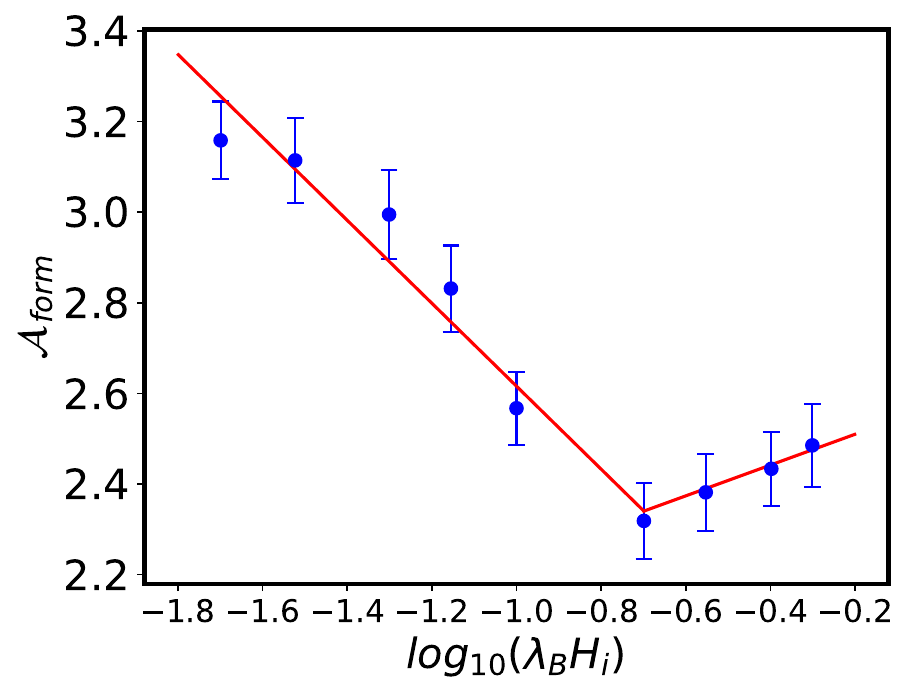}
\end{subfigure}
\
\begin{subfigure}{0.4\textwidth}
\includegraphics[width=\textwidth]{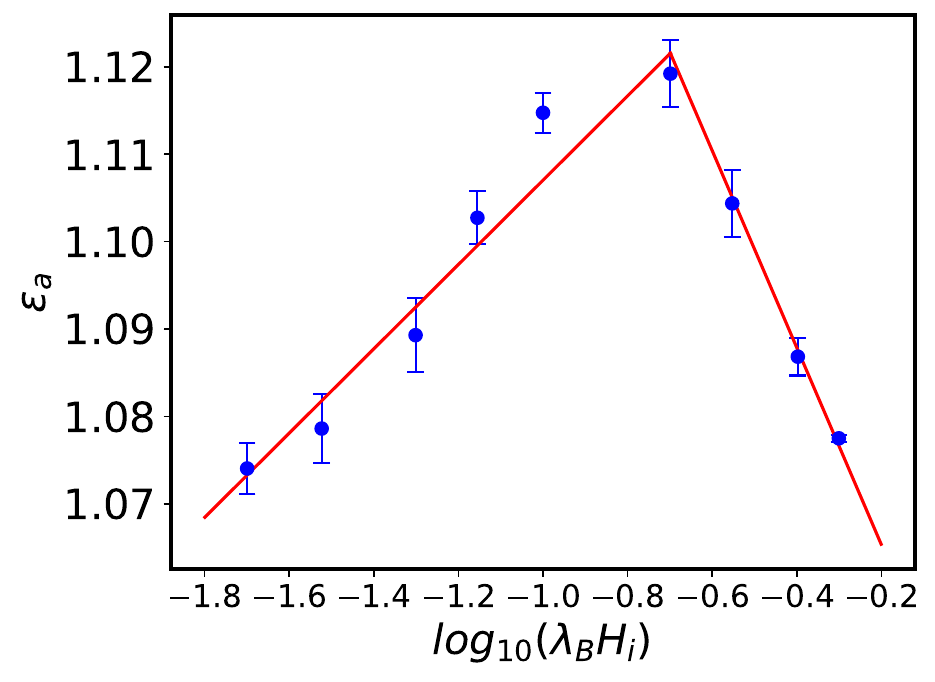}
\end{subfigure}
\begin{subfigure}{0.4\textwidth}
\includegraphics[width=\textwidth]{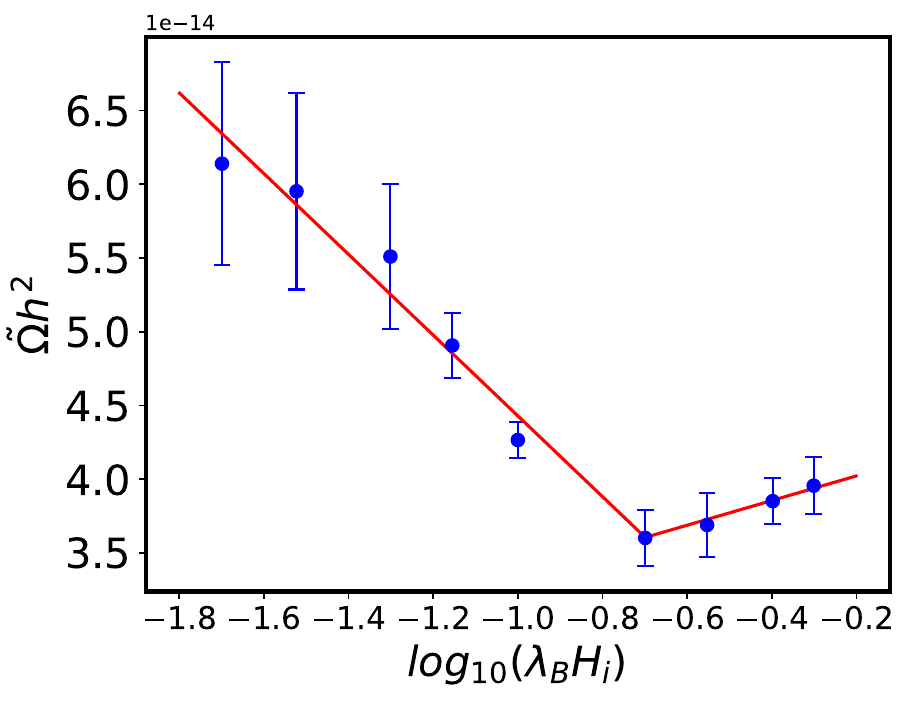}
\end{subfigure}
\caption{These results are all from the case $\sigma_M=1$ for different $\lambda_B$. The red curve is the fitting function.}
\label{fig:results_lambda(10.30)}
\end{figure}

\noindent{\it \bfseries The results of fitted parameters.} We first present the parameters $\log_{10}(\frac{t_{\text{dec}}}{t_{\text{form}}}),\mathcal{A}_{\text{form}}$, $\epsilon_a$ and $\tilde{\Omega}$ for different $\lambda_B$ with fixed $\sigma_M = 1$ in Fig.\ref{fig:results_lambda(10.30)}. All of these parameters can be fitted using piecewise linear or linear functions, which reflect the differing influences of long and short correlation lengths, as discussed in the main text. We perform five simulations for each case, and the error bars shown in these figures represent the numerical uncertainties of the extracted parameters. The fitted trends remain stable within these uncertainties, indicating the robustness of the results. Using these results and the corresponding fitting functions, we can obtain the expressions in Eq.11 and Eq.14 in the main text. We here note that, the momentum parameter in our simulations, $\epsilon_a \sim 1.1$, is smaller than the value $\epsilon_a \sim 1.85$ reported in Ref.\cite{1412.0789}, while our scaling parameter at formation, $\mathcal{A}_{\text{form}} \sim 3$, is larger than their reported value $\mathcal{A}\sim 0.8$ for $N_{\rm DW}=3$. Moreover, Ref.\cite{1412.0789} assumes that the DW network remains in a scaling regime, whereas in our simulations the network radiates continuously. We therefore assume that the entire DW energy density at formation is converted into axion, resulting in a larger axion emission than that obtained in Ref.\cite{1412.0789}. 

\begin{figure}[!htp]
\centering
\begin{subfigure}{0.4\textwidth}
\includegraphics[width=\textwidth]{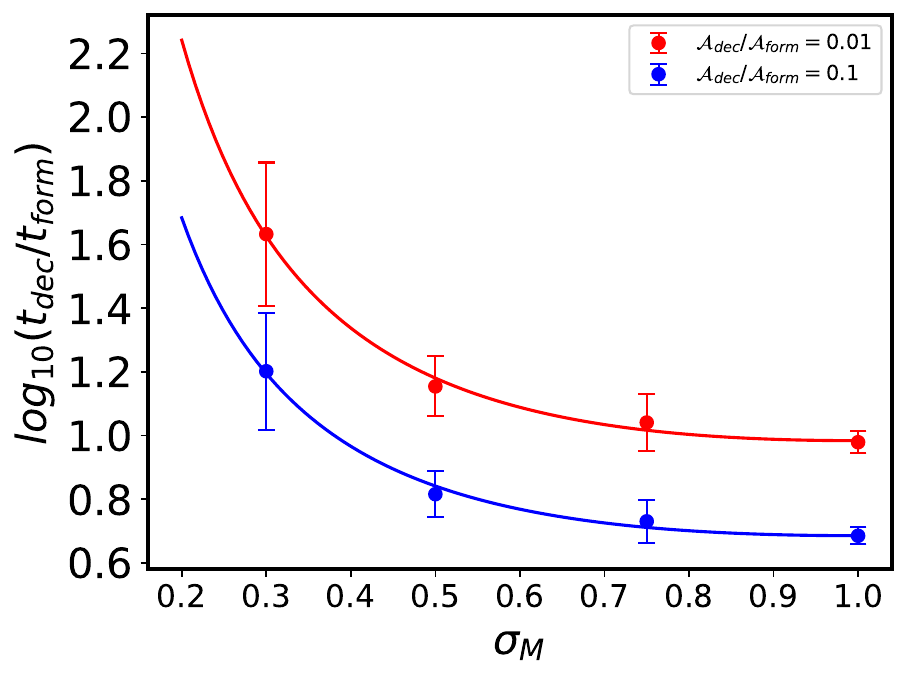}
\end{subfigure}
\begin{subfigure}{0.4\textwidth}
\includegraphics[width=\textwidth]{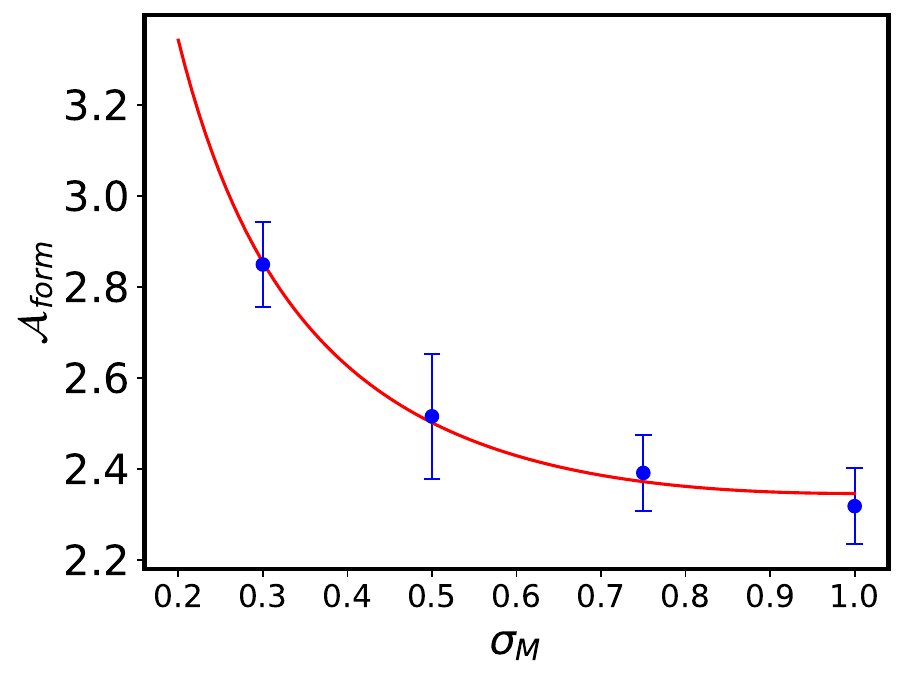}
\end{subfigure}
\
\begin{subfigure}{0.4\textwidth}
\includegraphics[width=\textwidth]{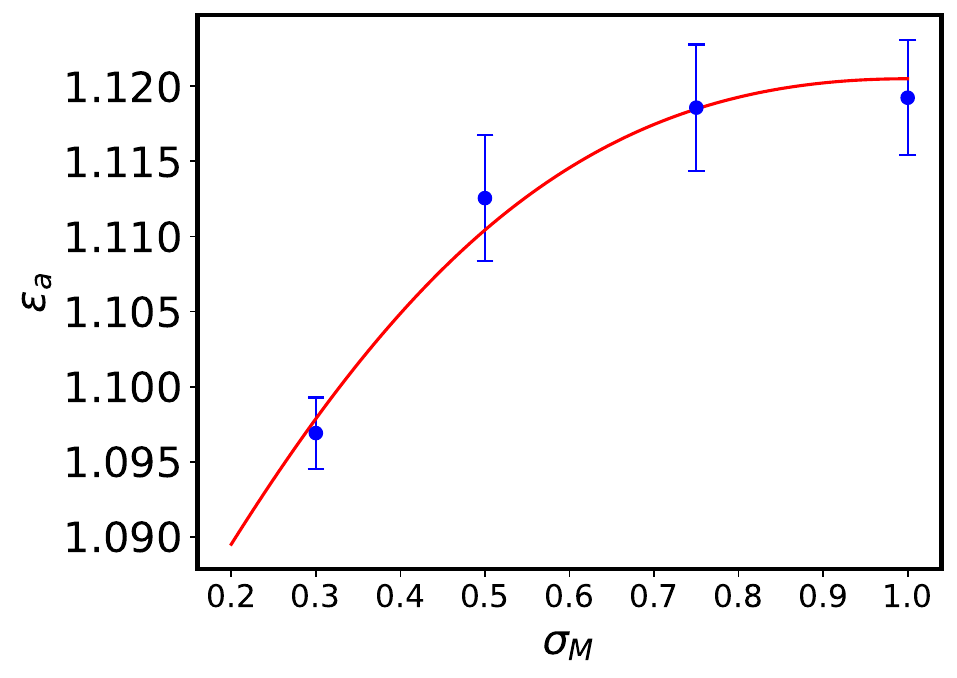}
\end{subfigure}
\begin{subfigure}{0.4\textwidth}
\includegraphics[width=\textwidth]{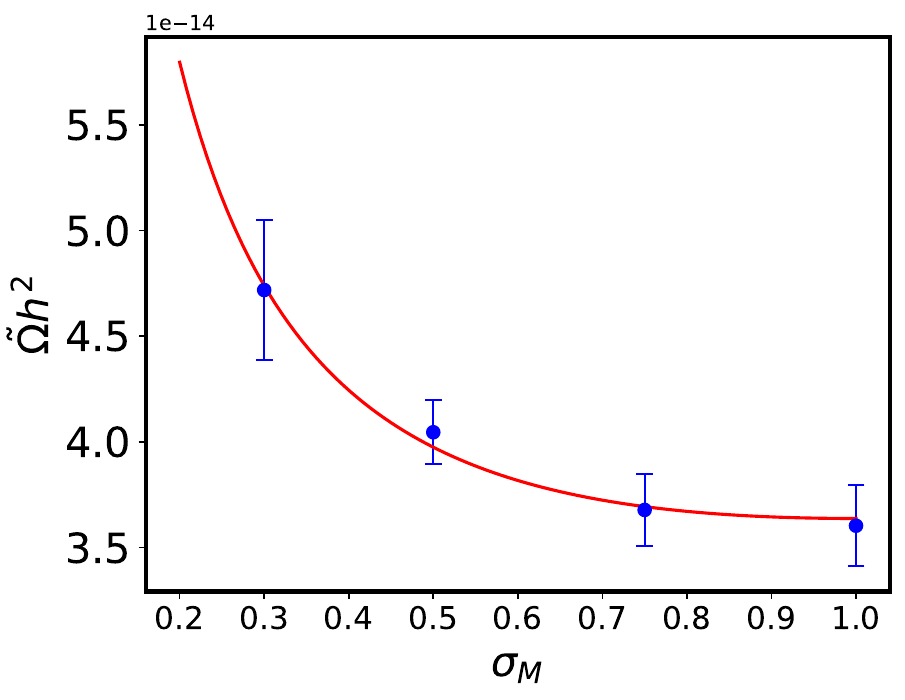}
\end{subfigure}
\caption{These results correspond to $\lambda_B\sim0.2H_i^{-1}$ with varying $\sigma_M$. The red curve is the fitting function.}
\label{fig:results_sigma(10.30)}
\end{figure}

In addition, we show the results of these parameters for different values of $\sigma_M$ with $\lambda_B \sim 0.2 H_i^{-1}$ in Fig.~\ref{fig:results_sigma(10.30)}. The ratio $\log_{10}(\frac{t_{\text{dec}}}{t_{\text{form}}})$ is proportional to $1/f_H$ and the scaling parameter at the formation of the DW network, $\mathcal{A}_{\text{form}}$, is also well fitted by $\mathcal{A}_{\text{form}} \propto 1/f_H$, which is consistent with the interpretation that faster-decaying DW networks form with smaller areas. The average momentum parameter $\epsilon_a$ follows $\epsilon_a \propto f_H$, indicating that faster decay produces higher-frequency axions for the same $\lambda_B$. Finally, the peak amplitude of the GW spectrum, $\Omega_{g w,0}^{p} h^2$, satisfies $\Omega_{gw,0}^{p} h^2 \propto 1/f_H$, in agreement with the fact that rapidly decaying DW networks emit less GW energy.

To briefly test the resolution and finite-volume stability, we have performed an additional check for a representative case, $\lambda_B = 0.2H_i^{-1}$ and $\sigma_M = 1$, using different lattice resolution and box size. The DW evolution shows no significant qualitative change among these runs. Since the evolution of the scaling parameter $\mathcal{A}$ captures the essential decay mechanism induced by the CS term, the agreement suggests that the main physical picture is stable against these numerical choices in the parameter range relevant to our analysis.

\begin{figure}[!htp]
\centering
\begin{subfigure}{0.4\textwidth}
\includegraphics[width=\textwidth]{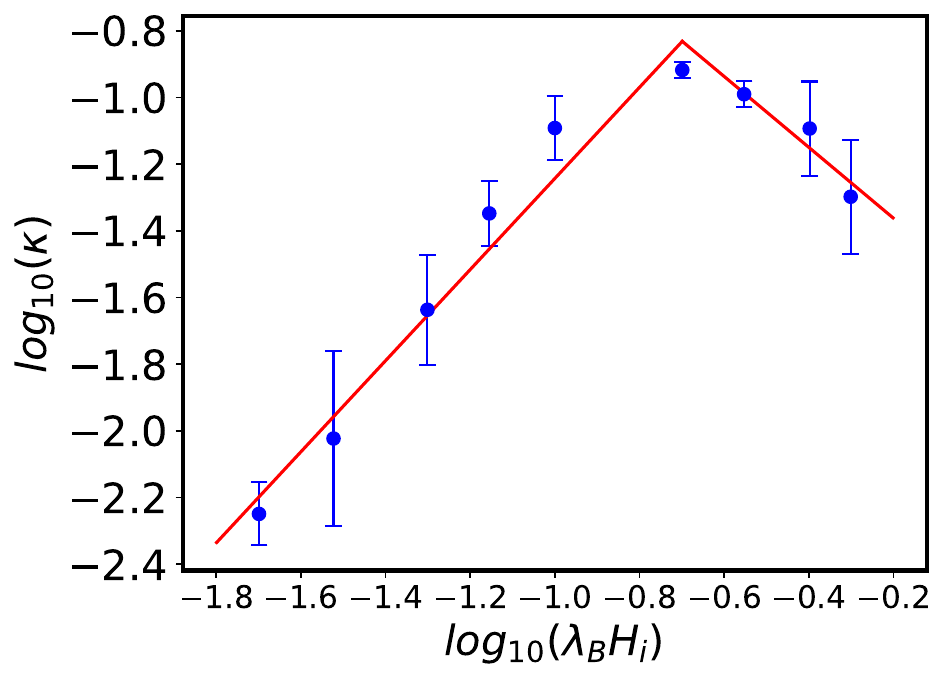}
\end{subfigure}
\begin{subfigure}{0.4\textwidth}
\includegraphics[width=\textwidth]{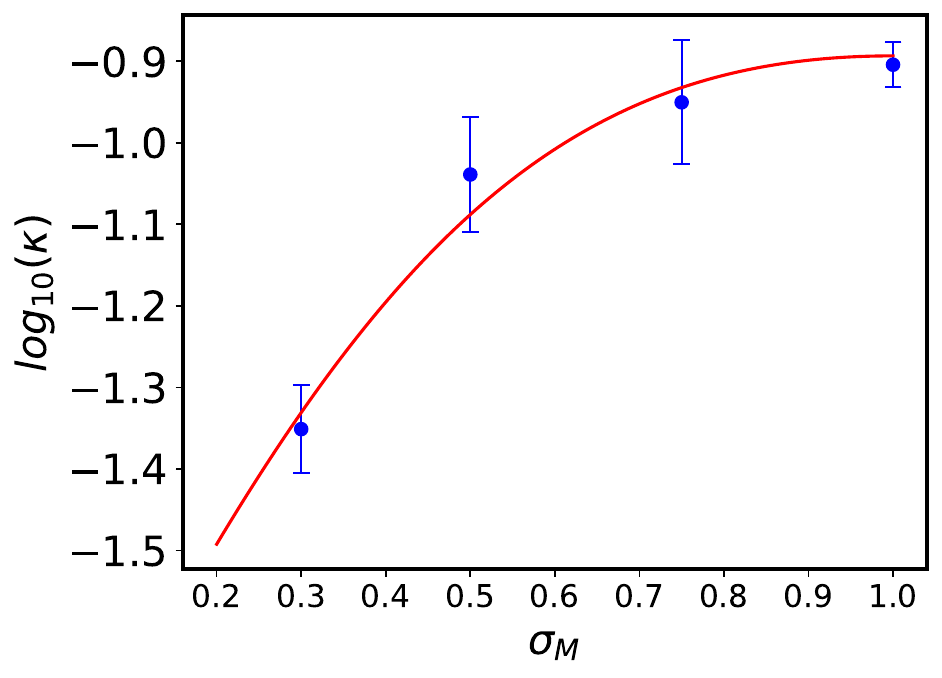}
\end{subfigure}
\caption{The efficiency factor of axion radiation $\kappa$. Left: The cases for different $\lambda_B$ and fixed $\sigma_M=1$. Right: The cases for different $\sigma_M$ and fixed $\lambda_B\sim 0.2H^{-1}_i$.}
\label{fig:kappa(10.30)}
\end{figure}

For different decay times and different time-dependent behaviors $p(t)$, we can obtain the efficiency factor $\kappa$ for different cases. We plot the $\lambda_B$ and $\sigma_M$ dependence of $\kappa$ in Fig~\ref{fig:kappa(10.30)}.

\begin{figure}
\centering
\begin{subfigure}{0.4\textwidth}
\includegraphics[width=\textwidth]{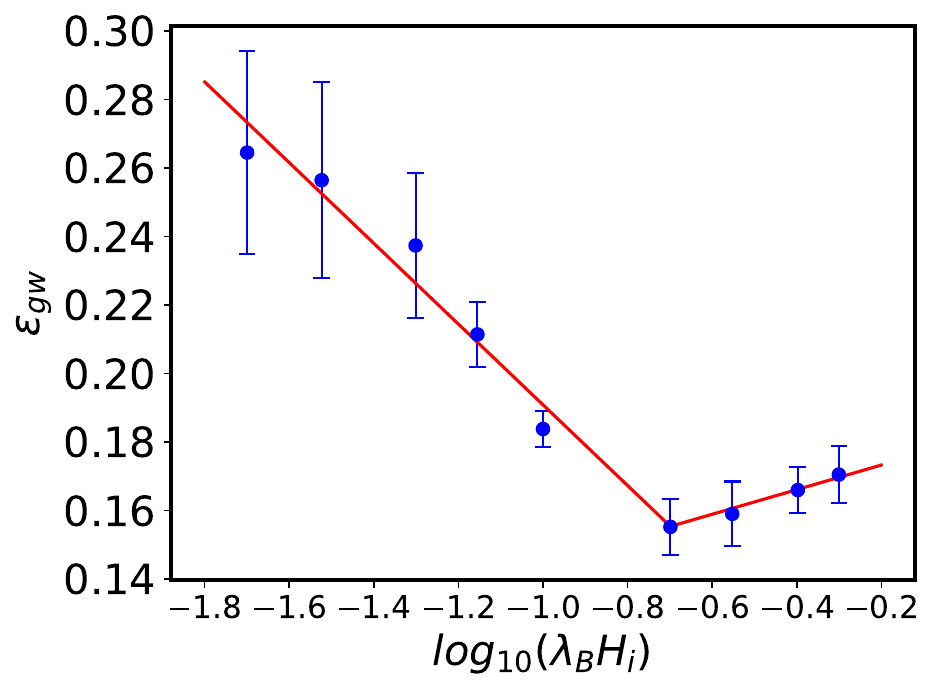}
\end{subfigure}
\begin{subfigure}{0.4\textwidth}
\includegraphics[width=\textwidth]{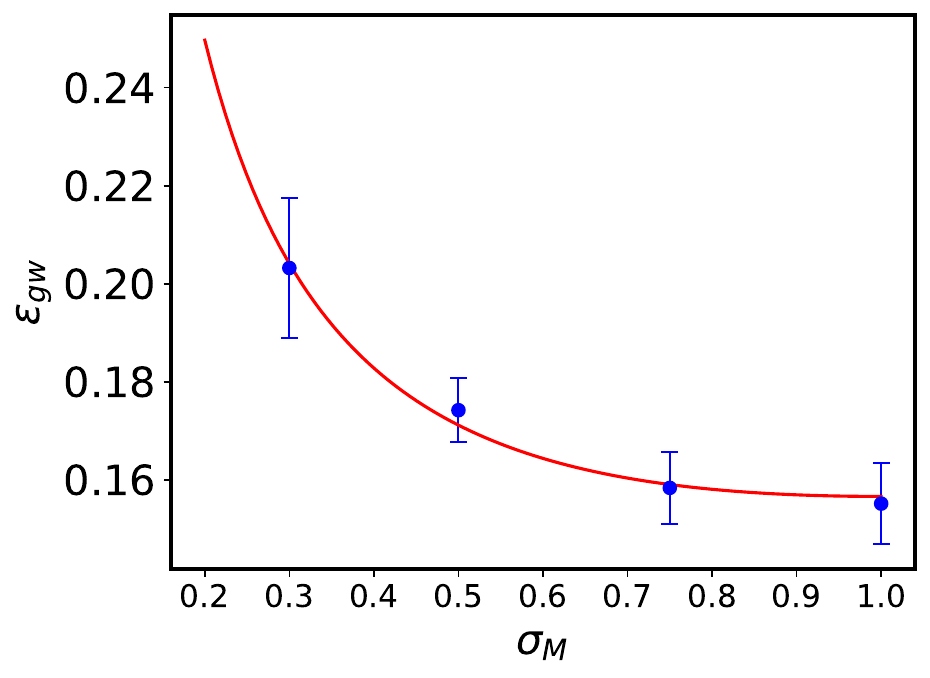}
\end{subfigure}
\caption{The efficiency parameter of GW radiation $\epsilon_{gw}$. Left: The cases for different $\lambda_B$ and fixed $\sigma_M=1$. Right: The cases for different $\sigma_M$ and fixed $\lambda_B\sim 0.2H^{-1}_i$.}
\label{fig:epsilon_GW(10.30)}
\end{figure}

We then present the efficiency parameter of GW radiation. Rewriting the $\Omega_{gw}$ as
\begin{equation}
\label{eq:rewriting_Omega}
\Omega_{gw}=\frac{\rho_{gw}}{\rho_c}=\epsilon_{gw}\frac{8\pi G^2\sigma_{dw}^2}{3H^2}=\epsilon_{gw}\beta^2\frac{8\pi G^2m_a^2f_a^4}{3H^2},
\end{equation}
where we define the efficiency parameter $\epsilon_{gw}\equiv \rho_{gw}/(G\sigma_{dw}^2)$. Using the fitted $\Omega_{gw}^p$, we obtain $\epsilon_{gw}$ for each case, as shown in Fig~\ref{fig:epsilon_GW(10.30)}.

For the quantities $\epsilon_a, \mathcal{A}_{\text{form}}, \tilde{\Omega}$, and the decay-time ratio $t_{\text{dec}}/t_{\text{form}}$, we have already derived their dependence on the PMF correlation length $\lambda_B$. We now extend this analysis to determine how these quantities scale with the magnitude of the gauge–axion coupling term, $C=\alpha F_{\mu\nu}\tilde{F}^{\mu\nu}$. Based on the discussion in the text, this coupling term can be parameterized as
\begin{equation}
\label{eq:para_coupling}
C(\alpha, B_{\text{ini}}, f_H)=\left(\frac{\alpha}{\alpha_0}\right)\left(\frac{B_{ini}^2}{B_0^2}\right)\left(\frac{f_H}{f_{H,0}}\right) C_0,
\end{equation}
where $\alpha_0=0.2, f_{H,0}=1$ are the reference values adopted in our simulations. In these simulations, we varied $f_H$ while keeping $\alpha=\alpha_0, B_{\mathrm{ini}}=B_0$ for $\lambda_B\sim 0.2H^{-1}i$ case and obtained the dependence of $\epsilon_a, \mathcal{A}_{\text{form}}, \tilde{\Omega}$, and $t_{\text{dec}}/t_{\text{form}}$ on $f_H$. Taking $\mathcal{A}_{\text{form}}$ as an example, we found $\mathcal{A}_{\text{form}} \propto \frac{1}{f_H}$. Since the physical quantity that governs the dynamics is the magnitude of the coupling term $C$ rather than $f_H$ alone, the above dependence can be rewritten as

\begin{align}
\label{eq:example_Aform_1}
&\mathcal{A}_{\text{form}}(\alpha_0, B_0, f_H)=\mathcal{A}_{\text{form}}(\alpha_0, B_0, f_{H,0})+k_f\left(\frac{1}{f_H}-\frac{1}{f_{H,0}}\right)=\mathcal{A}_{\text{form}}(\alpha_0, B_0, f_{H,0})+\frac{k_f}{f_{H,0}}\left(\frac{C_0}{C(\alpha_0, B_0, f_H)}-1\right), \notag \\
&\mathcal{A}_{\text{form}}(\alpha, B_0, f_{H,0})=\mathcal{A}_{\text{form}}(\alpha_0, B_0, f_{H,0})+k_{\alpha}\left(\frac{1}{\alpha}-\frac{1}{\alpha_0}\right)=\mathcal{A}_{\text{form}}(\alpha_0, B_0, f_{H,0})+\frac{k_{\alpha}}{\alpha_0}\left(\frac{C_0}{C(\alpha, B_0, f_{H,0})}-1\right).
\end{align}
If we choose parameters satisfying $f_{H,1}/f_{H,0} = \alpha_1/\alpha_0$, then $C(\alpha_0,B_0,f_{H,1}) = C(\alpha_1,B_0,f_{H,0})$, which implies $\mathcal{A}_{\text{form}}(\alpha_0,B_0,f_{H,1}) = \mathcal{A}_{\text{form}}(\alpha_1,B_0,f_{H,0})$. This equality requires
$\frac{k_f}{f_{H,0}} = \frac{k_{\alpha}}{\alpha_0}$. An analogous argument applies to variations in $B_{\mathrm{ini}}$. Combining all contributions, the general dependence of $\mathcal{A}_{\text{form}}$ on the couplings $(\alpha, B_{\mathrm{ini}}, f_H)$ can be expressed as
\begin{equation}
\mathcal{A}_{\text{form}}(\alpha, B_{ini}, f_{H})=\mathcal{A}_{\text{form}}(\alpha_0, B_0, f_{H,0})+k\left(\frac{\alpha_0}{\alpha}\frac{B^2_0}{B_{ini}^2}\frac{f_{H,0}}{f_H}-1\right),
\end{equation}
where $k=k_f/f_{H,0}$ is determined from simulation. The same procedure applies to the other parameters $\epsilon_a, \tilde{\Omega}$ and $t_{\text{dec}}/t_{\text{form}}$. In our analysis, we assume that the slope $k$ remains approximately independent of the correlation length $\lambda_B$.

\noindent{\it \bfseries A conservative estimate of the free-streaming scale.} In general, axion production associated with primordial magnetic fields can be constrained by free streaming~\cite{Kamada_2017}. We now estimate the free-streaming scale of the axion in order to examine whether the axion radiated by the DW network behave as sufficiently cold dark matter. As shown in Ref.~\cite{free-streaming_1,free-streaming_2}, the maximal comoving free-streaming length of dark matter with a characteristic momentum $q_*$ is given by
\begin{equation}
\lambda_{FS}=\frac{\sqrt{2}}{a_{eq}H_{eq}}\frac{q_*}{a_{eq}m_a}\ln\left( \frac{8a_{eq}m_a}{q_*} \right).
\end{equation}
In our simulation, the axion spectrum is broad, and it is therefore difficult to identify a unique characteristic momentum. For this reason, we instead estimate the free-streaming length using the maximal momentum $k_{max}$ at the end of the simulation, which is still relativistic and satisfies $k_{\max}\simeq 4m_a$. The scale factor at which this mode becomes nonrelativistic is then $a_{GM}=a_{end}k_{max}/m_a$, where $a_{end}$ is the scale factor at the end of the simulation, and $a_{end}k_{max}$ is the corresponding maximal comoving momentum. With the above formula, we have,
\begin{equation}
\lambda_{FS}=\frac{\sqrt{2}}{a_{eq}H_{eq}}\frac{a_{GM}}{a_{eq}}\ln\left( \frac{8a_{eq}m_a}{q_*} \right),
\end{equation}
and obtain
\begin{equation}
\lambda_{FS}\sim 0.00018~\text{Mpc},
\end{equation}
which is much smaller than the constraint $\lambda_{FS}\sim 0.084~\text{Mpc}$ quoted in Ref.\cite{free-streaming_1}.

This estimate is deliberately conservative, since it assigns the largest momentum present in the simulated spectrum to the entire axion population. In reality, the axion momentum distribution contains a substantial fraction of lower-momentum modes, and its true free-streaming length should therefore be smaller than the value obtained above. Hence, the fact that the bound $\lambda_{FS}\sim 0.084~\text{Mpc}$ is satisfied even when using $k_{max}$ indicates that the emitted axions are safely consistent with cold-dark-matter behavior from the perspective of free streaming.

\begin{figure}
\centering
\begin{subfigure}{0.5\textwidth}
\includegraphics[width=\textwidth]{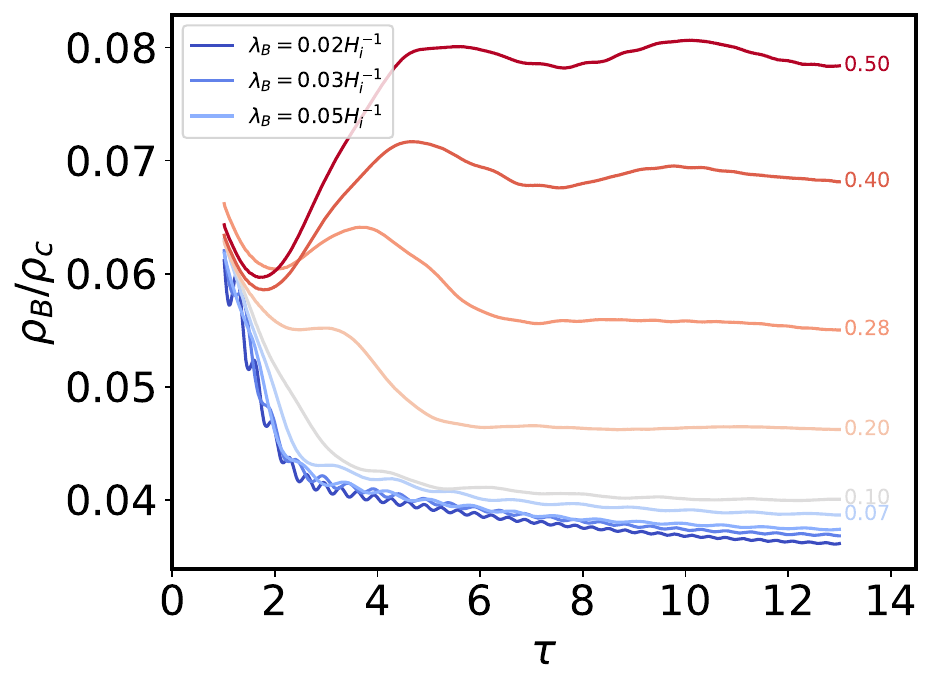}
\end{subfigure}
\caption{This figure shows the evolution of the ratio $\rho_B/\rho_c$ for each case, where the value at the tail of each curve indicates the corresponding $\lambda_B H_i$ associated with that case. The black curve corresponds to the case $\alpha = 0$. }
\label{fig:magnetic_field_energy(10.30)}
\end{figure}

\noindent{\it \bfseries The backreaction of magnetic energy density} We now consider the evolution of magnetic field energy density, which can show the backreaction of the gauge-coupling. The ratio of the magnetic field energy density to the background radiation energy density $\rho_B/\rho_c$ is plotted in  Fig.~\ref{fig:magnetic_field_energy(10.30)}. For the case without coupling $\alpha=0$, this ratio decreases as the number of relativistic degrees of freedom $g_*(T)$ decreases. When $\alpha \neq 0$, the coupling transfers energy from the axion field to the gauge field, causing the magnetic field energy density to increase. From this figure, we also observe that the energy growth is slightly larger for magnetic fields with longer correlation lengths; however, this enhancement is too small to raise the magnetic field amplitude above the observational bounds.

\begin{figure}
\centering
\begin{subfigure}{0.3\textwidth}
\includegraphics[width=\textwidth]{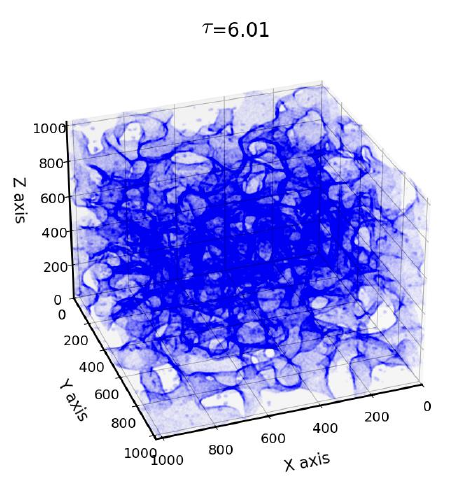}
\end{subfigure}
\begin{subfigure}{0.3\textwidth}
\includegraphics[width=\textwidth]{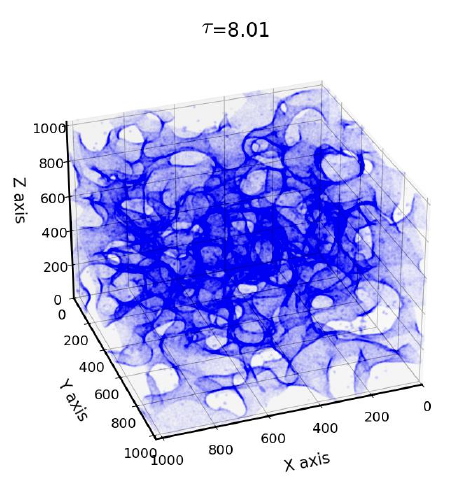}
\end{subfigure}
\begin{subfigure}{0.3\textwidth}
\includegraphics[width=\textwidth]{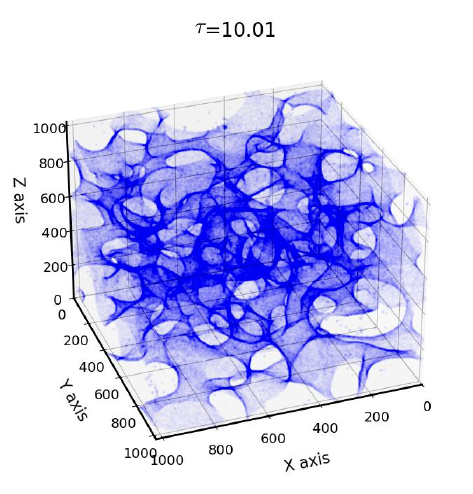}
\end{subfigure}
\
\begin{subfigure}{0.3\textwidth}
\includegraphics[width=\textwidth]{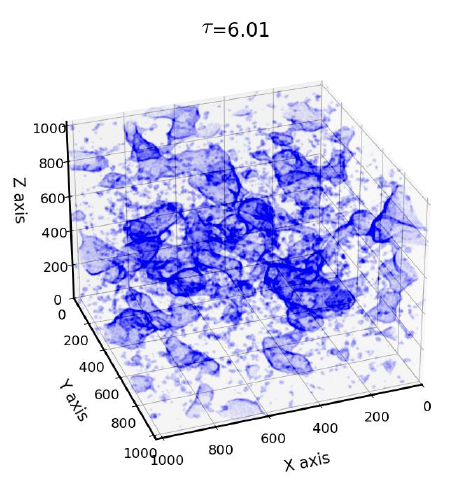}
\end{subfigure}
\begin{subfigure}{0.3\textwidth}
\includegraphics[width=\textwidth]{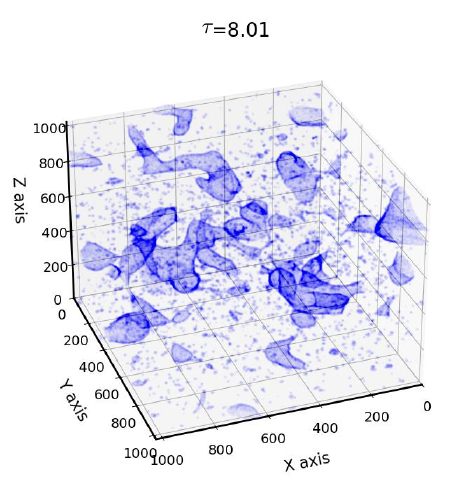}
\end{subfigure}
\begin{subfigure}{0.3\textwidth}
\includegraphics[width=\textwidth]{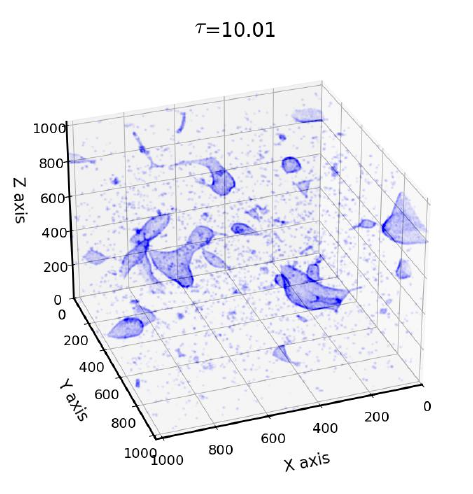}
\end{subfigure}
\
\begin{subfigure}{0.3\textwidth}
\includegraphics[width=\textwidth]{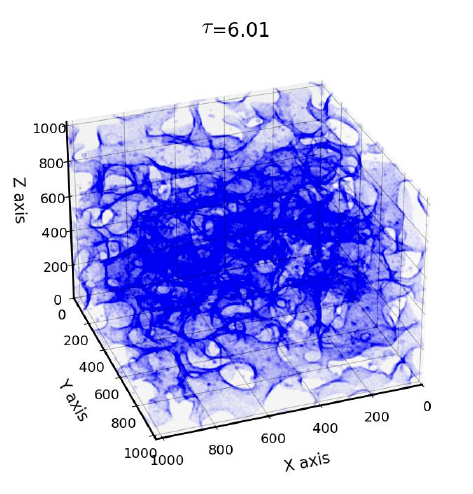}
\end{subfigure}
\begin{subfigure}{0.3\textwidth}
\includegraphics[width=\textwidth]{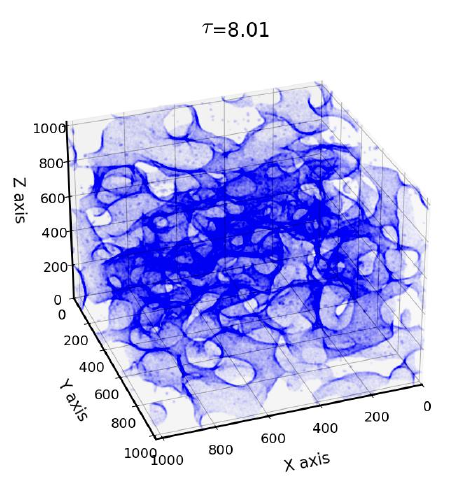}
\end{subfigure}
\begin{subfigure}{0.3\textwidth}
\includegraphics[width=\textwidth]{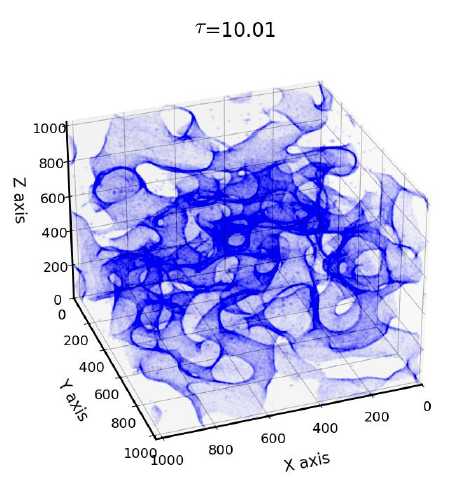}
\end{subfigure}
\

\caption{The distribution of DW network at time $\tau=6.01$, $\tau=8.01$ and $\tau=10.01$. From top to bottom, the figures show the results of case $\alpha=0$, $\lambda_B=0.1H_i^{-1}$, and $\lambda_B=0.02H_i^{-1}$.}
\label{fig:DW_network(09.17)}

\end{figure}

\noindent{\it \bfseries 3D distribution of the DW network.} We plot the distribution of the DW network for each case at the same time in Fig.~\ref{fig:DW_network(09.17)}. These figures provide an intuitive illustration of the decay behavior of DW networks coupled to helical PMFs and show the different decay rates for different PMF correlation lengths.

\end{document}